\documentclass[12pt,pdftex]{article}

\pdfoutput=1

\usepackage[a4paper,text={16.8cm,22.4cm}]{geometry}
\usepackage{amsmath,amssymb,bm,graphicx,color}
\allowdisplaybreaks 
\newcommand\slurp[1]{#1}
\newcommand\sslurp[2]{#1{#2}}
{\catcode`/=\active \expandafter}%
\slurp{\newcommand/}{/\penalty1000\hskip0pt\relax}
{\catcode`:=\active \expandafter}%
\slurp{\newcommand:}{:\penalty1000\hskip0pt\relax}

\newcommand\addspace{\ifcat\nextchar a\spacefactor999. \else.\fi}
{\catcode`\.=\active \expandafter}%
\slurp{\newcommand.}{\futurelet\nextchar\addspace}

\usepackage[unicode]{hyperref} 
\ifx\href\undefined\def\href#1{}\fi
\ifx\texorpdfstring\undefined\def\texorpdfstring#1#2{#1}\fi

\newcommand\myslash{/} \newcommand\mycolon{:}
\newcommand\doi{{\catcode`/=\active \catcode`:=\active \expandafter}\sslurp\realdoi}
{\catcode`/=\active \catcode`:=\active \expandafter}%
\slurp{\newcommand\realdoi[1]{{\let/=\myslash \let:=\mycolon
                               \edef\raw{{http://dx.doi.org/#1}}\expandafter}%
                               \expandafter\href\raw{doi:#1}}}
\newcommand\eprint[2]{{\escapechar-1%
                       \edef\a{\expandafter\string\csname arXiv\endcsname}%
                       \edef\b{\expandafter\string\csname #1\endcsname}%
                       \edef\c{\expandafter\string\csname #2\endcsname}%
                       \edef\d{\noexpand\href{http://arXiv.org/abs/\c}}%
                       \ifx\a\b\expandafter\d\fi{\tt #1:#2}}}
\newcommand{\be}{\begin{equation}}
\newcommand{\ee}{\end{equation}}

\def\d{{\rm d}}
\def\OMIT#1{{}}

\newcommand{\mcdot}{\!\cdot\!}

\newcommand{\bra}[1]{\left\langle #1\right\rvert}
\newcommand{\ket}[1]{\left\lvert #1\right\rangle}

\newcommand{\e}{\mathrm{e}}

\newcommand{\eq}[1]{Eq.~\eqref{#1}}

\newcommand{\vc}[1]{\boldsymbol{#1}}

\begin{document}
\begin{titlepage}

\begin{flushright}
% LAUR-??-??
\end{flushright}

\vspace{0.2cm}
\begin{center}
\LARGE\bf
WIMP-nucleus scattering \\  in chiral effective theory
\end{center}

\vspace{0.2cm}
\begin{center}
Vincenzo Cirigliano$^{a}$, Michael L. Graesser$^a$, and Grigory Ovanesyan$^a$

\vspace{0.6cm}
{\sl 
${}^a$\, Theoretical Division, Los Alamos National Laboratory \\
MS B283, Los Alamos, NM 87545, U.S.A.\\[2mm]
}
\end{center}

\vspace{0.2cm}
\begin{abstract}
\vspace{0.2cm}
\noindent

We discuss  long-distance QCD corrections to the WIMP-nucleon(s) interactions 
in the framework of chiral effective theory. 
For  scalar-mediated WIMP-quark interactions, 
we calculate all the next-to-leading-order corrections to the WIMP-nucleus elastic cross-section,
including two-nucleon amplitudes and recoil-energy dependent shifts to the single-nucleon scalar form factors. 
As a consequence, 
the scalar-mediated WIMP-nucleus  cross-section cannot be parameterized 
in terms of just two quantities, namely the neutron and proton scalar form factors  
at zero momentum transfer, but additional parameters appear, depending  on the short-distance 
WIMP-quark interaction. Moreover, multiplicative factorization of the cross-section into  particle, nuclear and astro-particle parts is violated. 
In practice,  while the new effects are of the  natural size expected by chiral power counting, 
they become  very important  in those regions of parameter space where  
the leading order WIMP-nucleus amplitude  
is  suppressed,  including the so-called  ``isospin-violating dark matter"  regime. 
In these regions of parameter space  we find order-of-magnitude corrections to the total scattering rates 
and qualitative changes to the shape of recoil spectra.

\end{abstract}
\vfil

\end{titlepage}

\section{Introduction}\label{sec:introduction}
The evidence for Dark Matter (DM) is overwhelming. However currently all experimental evidence for its existence comes 
from astrophysical observations. Great progress has been made in last few years in laboratory underground experiments trying to measure elastic recoil of  WIMPs (weakly interacting massive particles) from a nucleus inside the detector. 
The positive results reported by DAMA \cite{Bernabei:2008yi}  and CoGeNT \cite{Aalseth:2011wp} are in contradiction with null results from XENON  \cite{Aprile:2011hi, Angle:2011th}and CDMS \cite{Ahmed:2012vq, Ahmed:2010wy}, 
when analyzed in a standard WIMP scenario~\cite{Farina:2011pw,Kopp:2011yr}, and 
many ideas have been proposed to resolve this puzzle. 
These ideas mostly involve modifications to the short-distance dynamics,  
such as the inelastic dark matter scenario~\cite{TuckerSmith:2001hy,Graham:2010ca,Cline:2010kv}, 
isospin violating dark matter couplings  (IVDM)~\cite{Kurylov:2003ra,Giuliani:2005my,Chang:2010yk,Feng:2011vu}, 
momentum-dependent couplings~\cite{Feldstein:2009tr,Chang:2009yt}, 
and resonant scattering~\cite{Bai:2009cd}. 
The impact of astrophysical uncertainties has also been explored in the recent 
literature~\cite{Savage:2006qr,Lisanti:2010qx,Fox:2010bz,Frandsen:2011gi,HerreroGarcia:2012fu}.

Here we take a complementary  point of view. We wish to study how 
known hadronic and nuclear physics effects (such as nucleon form factors and 
meson exchange currents)  affect  WIMP-nucleus elastic scattering, 
and explore whether the inclusion of these effects  helps mitigate the apparent contradictions between 
DAMA, CoGeNT and XENON and CDMS. 
The need to revisit the hadronic and nuclear physics of WIMP-nucleus scattering 
has been pointed out in the recent literature, 
and first systematic studies in this direction have appeared in  Refs.~\cite{Fan:2010gt,Fitzpatrick:2012ix}. 
Ref.~\cite{Fan:2010gt}   developed  a non-relativistic effective field  theory  (EFT) in which the degrees of freedom are 
the WIMP and nucleus as a whole, and studied the set of distinguishable recoil spectra that could arise 
from different underlying models.
The authors of Ref.~\cite{Fitzpatrick:2012ix}, on the other hand,   focused on a non-relativistic EFT 
at the WIMP-nucleon level,  wrote the most general set of  single-nucleon operators  up to second order in 
momentum transfer,  and worked out the corresponding nuclear responses (for relevant targets)  
within the nuclear shell model. 

Our current work adds yet a different spin, in that it discusses  the WIMP-nucleon 
interaction within the  chiral EFT  framework, which incorporates 
at the nucleon level the consequences of the broken chiral symmetry of QCD. 
For simplicity we discuss here only the case of  scalar-mediated  WIMP-quark interaction 
and derive the leading  order (LO)  and next-to-leading order (NLO) WIMP-nucleon operators. 
To NLO,   the resulting WIMP-nucleon operators involve 
not only the single nucleon (scalar) form factor,   but also 
a two-nucleon operator,  generated by a so-called meson-exchange diagram. 
The latter effect has been previously considered in  Ref. \cite{Prezeau:2003sv}, 
where it was  claimed it could affect WIMP-nucleus cross-sections at the $\sim 50 \%$ level. 
Compared to  the analysis of Ref.~\cite{Prezeau:2003sv},  we embed the meson-exchange diagram 
in a consistent chiral $SU(3)$  power counting to NLO,  that includes  $\eta-\eta$ exchange in addition to $\pi-\pi$ exchange
and also loop corrections to the one-nucleon amplitudes (not considered in  \cite{Prezeau:2003sv}).
Compared to Ref. \cite{Fitzpatrick:2012ix} our work considers a more restricted underlying interaction
by assuming that WIMP  interacts with quarks via the scalar density.
Within this restricted setting, however,
we go beyond Ref.~\cite{Fitzpatrick:2012ix}  in several respects: 
(i) Working to NLO in the ratio of momentum transfer to nucleon mass, 
we include all the interactions consistent with QCD. 
These include form-factor corrections to the one-nucleon operators
(which are left unspecified in  Ref.~\cite{Fitzpatrick:2012ix}) 
as well as  two-nucleon operators not considered in Ref.~\cite{Fitzpatrick:2012ix}.
To our knowledge this is the first time the chiral power  corrections are consistently included for WIMP-nucleus elastic cross-section.
(ii)   Because we start from the underlying WIMP-quark interaction, 
we are able to relate the  coefficients of the WIMP-nucleon operators 
(left arbitrary in the bottom-up nucleon-level  EFT approach of  \cite{Fitzpatrick:2012ix})  
to the  short-distance parameters of the theory.

The paper is organized as follows. 
In Section \ref{sec:scalar} we set up the framework for our discussion, 
specifying the short-distance  scalar-mediated WIMP-quark interaction in a model-independent way. 
In Section \ref{sec:chpt} we  discuss the chiral perturbation theory (ChPT) framework and 
identify the LO  and NLO graphs that contribute to scalar-mediated  WIMP scattering off  nucleons. In Section
\ref{sect:nlo} we compute the relevant NLO corrections, that include  one loop diagrams (\ref{sec:loops})
two-body meson-exchange  diagrams (\ref{sec:TwoBodyOperators}). 
In Section \ref{sec:NSM} we discuss  the matrix elements of WIMP-nucleon operators in the nucleus, 
 and briefly  review the nuclear shell model (NSM) that we use to evaluate the matrix elements of two-body operators. 
We consider the  phenomenological implications  of our NLO corrections in Section \ref{sec:phenomenology}, 
and conclude  in section~\ref{sec:conclusions}.

\section{Scalar-mediated WIMP-quark interaction}\label{sec:scalar}

We consider the following model-independent interaction between WIMP and quarks:
\begin{eqnarray}
\mathcal{L}_{\chi q}=\sum_{q=u,d,s,c,b,t} \,k_q\,m_q\,\bar{q}q+k_G\,\frac{\alpha_s}{\pi} G_{\mu\nu}G^{\mu\nu},\label{EFTLag}
\label{eq:lag1}
\end{eqnarray}
where $k_{q,G}$ are functions of the DM field(s).  For example, in the case of
Dirac or Majorana fermion DM  one has  $k_{q,G} \propto \bar{\chi} \chi$.%
Integrating out the heavy quarks $c,b,t$~\cite{Shifman:1978zn} we get the following effective interaction:
\begin{eqnarray}
&&\mathcal{L}_{\text{eff}}=\sum_{q=u,d,s}\, s_q \bar{q}q+s_{\Theta}\,\Theta^{\mu}_{\mu},
\label{eq:lag2}
\end{eqnarray}
where we have used the trace anomaly equation and $\Theta^{\mu}_{\mu}$ is the trace of 
the stress energy tensor. 
The fields $s_q$ and $s_{\Theta}$ are related to $k_{q,G}$  as follows:
\begin{eqnarray}
s_{q}=m_q\left(k_q-\frac{2}{27}\sum_{Q=c,b,t}{k_Q}+\frac{8}{9} k_G\right), \qquad  s_{\Theta}=\frac{2}{27}\sum_{Q=c,b,t}{k_Q} - 
\frac{8}{9}  k_G~.
\label{eq:svsk}
\end{eqnarray}

Specializing to the case of fermionic DM particle (denoted by $\chi$),  we can  write 
the fields $k_{q,G}$ in terms of $\chi$ and dimensionless  short-distance parameters $\tilde \lambda_{q,G}$ 
\be
k_{q}  =   \frac{\tilde{\lambda}_{q}}{ v \, \Lambda^2_{\text{np}}} \   \bar{\chi}\chi  ~, 
\qquad  \qquad 
k_{G}  =   \frac{\tilde{\lambda}_{G}}{v \, \Lambda^2_{\text{np}}} \   \bar{\chi}\chi  ~, 
\ee
where $v = (\sqrt{2} G_F)^{-1/2}$  is the Higgs VEV and  $\Lambda_{\text{np}}$ denotes a generic new-physics scale. 
After integrating out the heavy quarks,  the low-energy effective Lagrangian  (\ref{eq:lag2})
contains only four dimensionless parameters  $\lambda_{u,d,s,\Theta}$:
\be
s_{q}  = \frac{m_q}{v}   \frac{\lambda_{q}}{\Lambda^2_{\text{np}}}      \   \bar{\chi}\chi  
\qquad \qquad 
s_{\Theta}  =  \frac{\lambda_{\Theta}}{v \, \Lambda^2_{\text{np}}}      \   \bar{\chi}\chi  ~. 
\label{eq:sqvslambdaq}
\ee
In our  phenomenological analysis we will use $\lambda_{u,d,s,\Theta}$,  but the reader should 
keep in mind that these are related to the short distance parameters by 
$ \lambda_q = \tilde{\lambda}_q  - (2/27) \sum_Q  \tilde{\lambda}_Q  + (8/9) \tilde{\lambda}_G $ 
and $\lambda_\Theta = (2/27) \sum_Q \tilde{\lambda}_Q  - (8/9) \tilde{\lambda}_G$. 

In order to compute the WIMP-nucleus  cross-sections from the short-distance Lagrangian of \eq{eq:lag1}
that couples WIMPs to quarks,  two steps are needed: 
(i) At an energy scale of the order of $\sim$ GeV
 one matches the WIMP-quark Lagrangian
 non-perturbatively  to a  WIMP-nucleon effective Lagrangian.   
For this purpose we use  ChPT~\cite{Weinberg:1978kz,Gasser:1983yg,Gasser:1984gg} 
(for reviews, see \cite{Bernard:1995dp} and \cite{Bedaque:2002mn}), 
which parameterizes the non-perturbative physics in a number of low-energy constants that can 
either be determined phenomenologically or computed  in lattice QCD~\cite{Kronfeld:2012uk}.
This step provides an effective potential describing the interaction of the  WIMP with nucleons.
(ii)  With the WIMP-nucleon interaction Hamiltonian at hand, one then calculates the 
amplitude for WIMP-nucleus elastic scattering. 
This step requires information on the  
wave-function of the target nucleus in the ground state. 
The WIMP-nucleus interaction can also be parameterized in terms of a non-relativistic effective Lagrangian~\cite{Fan:2010gt}.
Note that while for definiteness we are focusing here only on the case of scalar-mediated WIMP-quark interactions, 
a similar analysis can be performed for all types of quark-WIMP interactions (vector, axialvector, pseudoscalar, tensor) 
and will be presented elsewhere~\cite{Cirigliano:2012xx}.

\begin{figure}[t]
\begin{center}
\includegraphics[width=0.90\textwidth]{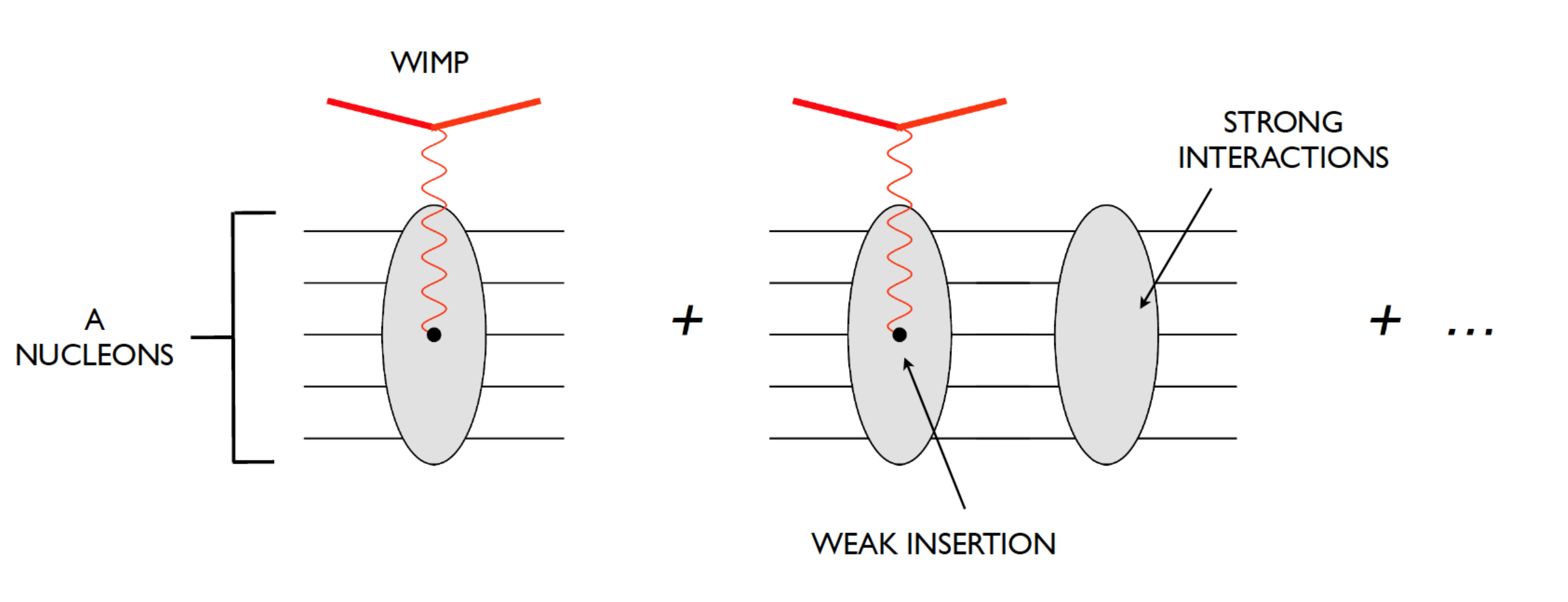}
% \parbox{15.5cm}
\caption{\label{fig:ladder} 
Representation of  ladder diagrams contributing to $T_{A,W}$,  
the scattering amplitude for $A$ nucleons and a WIMP. 
The ladder rungs are given by 
$A$-nucleon irreducible amplitudes; 
only one of the rungs  (denoted by $M_{A,W}$ in the text) 
involves a WIMP scalar density  insertion.}
\end{center}
\end{figure}

\section{WIMP-nucleon interactions in chiral effective theory}\label{sec:chpt}

Given the kinematics of WIMP-nucleus elastic scattering,  the three-momentum transfer 
to the hadronic system does not exceed $q_{\rm max} = 2  \mu_{WA} v_{\rm rel}  <   2  m_A v_0  \sim 200$~MeV, 
where $\mu_{WA}$ is the WIMP-nucleus reduced mass and 
we have used a typical target nucleus mass $m_A \sim 100$~GeV as well as  $v_0 \sim 10^{-3}$ for 
the  value at which the  velocity distribution of the dark matter halo starts to fall off exponentially.
Since $q_{\rm max}$ is small compared to the nucleon mass $m_N$, 
it is appropriate to  use ChPT to describe the WIMP-nucleon dynamics, 
expanding the amplitudes in $p \sim q/m_N \sim  m_{\pi,K,\eta}/m_N$. 

As suggested by the form of \eq{eq:lag2}, 
the weak interaction of WIMPs and light quarks 
can be incorporated in the framework of ChPT through the external source method~\cite{Gasser:1983yg},
i.e. adding to the QCD Lagrangian  external scalar sources $s(x) = {\rm diag} (s_u(x), s_d(x), s_s(x))$ 
coupled to the scalar quark density  
and the external source $s_\Theta (x)$ coupled  to the energy-momentum tensor.
Note that from the point of view of chiral symmetry, the external scalar  source transforms 
in the same way as the quark mass matrix $m_q$.
The octet of light pseudoscalar meson fields $(\phi_a)$  is described by the matrix $U = {\rm Exp} (i \sum_a  T_a \phi_a/F)$, 
where $F$ can be identified to leading order with the pion decay constant and  $T_a$ are the $SU(3)$ generators. 
In the chiral power counting one assigns the following scaling: $\partial U \sim O(p)$, 
while $m_q \sim s \sim  O(p^2)$ and $s_\Theta \sim O(p^0)$.  
In terms of these fields, the lowest order mesonic chiral Lagrangian reads 
\begin{eqnarray}
{\cal L}_M &= & {\cal L}_M^{(2)}  +   {\cal L}_M^{(4)}  +  \dots \qquad \qquad  {\cal L}_M^{(2n)} \sim O(p^{2n})~, 
\\
{\cal L}_M^{(2)}  &=&  \frac{F^2}{4} \, {\rm Tr} \Big[ \, \partial_\mu U^\dagger  \partial^\mu U \, \Big]  \ + \ 
\frac{ B_0 \, F^2}{2}  \, {\rm Tr} \Big[ \, (m_q  -   s) \, (U  + U^\dagger) \Big]~.
\end{eqnarray}
The low-energy constant (LEC) $B_0$ is related to the light quark condensate:   $\langle \bar{q} q \rangle = - F^2 B_0 (1 + O(m_q))$. 
For  the  meson-baryon Lagrangian we use the Heavy Baryon Chiral Perturbation Theory formulation~\cite{Jenkins:1990jv}, 
in which one writes  the baryon momentum as $p_B = m v + k$,  in terms of a velocity  $v_\mu$ and a residual momentum $k \sim O(p)$ 
and one  re-defines the  heavy baryon octet field $B$ such that $\partial B \sim O(p)$. 
Introducing the covariant derivative $\nabla_\mu B = \partial_\mu B  + [\Gamma_\mu, B]$ 
with $\Gamma_\mu =  1/2 ( u^\dagger \partial_\mu u  +  u \partial_\mu u^\dagger)$, 
the spin vector  $S_\mu = ( i/2) \gamma_5 \sigma_{\mu \nu} v^\nu$,   
the field  $u_\mu = i (u^\dagger \partial_\mu u -  u \partial_\mu u^\dagger)$ with $u = \sqrt{U}$,   
and finally  $\chi_+ = u^\dagger \chi u^\dagger  + u \chi^\dagger u$ with $\chi = 2 B_0  (m_q -  s)$, 
the first few terms in the chiral expansion 
read~\cite{Jenkins:1990jv,Bernard:1995dp} 
\begin{eqnarray}
{\cal L}_{MB}  &=&   {\cal L}_{MB}^{(1)}  + {\cal L}_{MB}^{(2)}  + \dots \qquad \qquad  {\cal L}_M^{(n)} \sim O(p^{n})~, 
 \\
{\cal L}_{MB}^{(1)}  &=&  {\rm Tr} \Big( \bar{B} \,  i v \cdot \nabla \, B \Big) + D \, {\rm Tr}  \Big( \bar{B} S^{\mu} \{u_\mu,B\} \Big)
 +  F \, {\rm Tr}  \Big( \bar{B} S^{\mu} [u_\mu,B]]\Big)
\label{eq:LMB1} 
 \\
{\cal L}_{MB}^{(2)}  &=&  b_D \, {\rm Tr}  \Big( \bar{B}  \{\chi_+ , B\} \Big) 
+ b_F \, {\rm Tr}  \Big( \bar{B}  [\chi_+ , B ]  \Big)
+ b_0 \, {\rm Tr}  \Big( \bar{B}   B \Big)  \, {\rm Tr} \Big( \chi_+  \Big)\label{eq:LMB2}
\end{eqnarray}
In the above expression $F,D,b_F,b_D,b_0$ are low-energy constants, related to the baryon
axial current matrix elements  ($F,D$), and the baryon mass splitting and sigma-terms 
($b_F,b_D,b_0$).

The EFT power counting allows one to identify the leading contributions to 
the scattering amplitude $T_{A,W}$
for the process    $N_1 + \dots + N_A + W  \ \longrightarrow \  N_1 + \dots + N_A + W $
involving  $A$ nucleons and a WIMP. 
As is well known~\cite{Weinberg:1990rz,Weinberg:1991um,Kaplan:1996xu},  when $A>1$ the 
naive  chiral power counting breaks down due to pinch singularities that arise when  
nucleons in the intermediate state simultaneously go on shell.  
As illustrated in Fig.~\ref{fig:ladder}, 
the full non-perturbative amplitude $T_{A,W}$ is obtained by summing a Lippmann-Schwinger series of ladder 
diagrams with $A$-nucleons intermediate states and rungs given by 
$A$-nucleon irreducible amplitudes (not necessarily connected), 
only one of which involves the insertion of the external probe (the WIMP scalar density in the case at hand).     
These $A$-nucleon irreducible amplitudes admit a consistent power counting, 
and the scaling of the full amplitude $T_{A,W}$ is controlled by the scaling of  $M_{A,W}$, 
the $A$-nucleon irreducible amplitude with insertion of the external probe. 

\begin{table}[!t]
\begin{center}
\begin{tabular}{|c|c|c|c|c|}
	\hline
Effective Lagrangian % $i$
& $d_i$&  $n_i$    &  $\epsilon_i\equiv d_i+\frac{n_i}{2}-2$      \\
	\hline
$\mathcal{L}_{M}^{(2n)}$& 2$n$ &  0 &   $2 (n-1)$ \\	
\hline
$\mathcal{L}_{MB}^{(n)}$ & $n$ & 2 & $ n-1 $  \\
	\hline
\end{tabular}
\caption{Chiral dimensions for vertices arising from  the purely mesonic and baryon-meson  effective Lagrangians. 
$n=1,2, \dots$ represents any positive integer.}
\label{tb1}
\end{center}
\end{table}

A diagram with $C$  connected parts, $L$ loops, $V_i$ strong-interactions  vertices of type $i$ 
and one  ``weak'' vertex  scales as $M_{A,W} \sim  p^{\nu}$ with~\cite{Weinberg:1991um,Bedaque:2002mn}
\be
\nu  = 4 - A  - 2 C  + 2 L  + \sum_{i} \,  V_i \, \epsilon_i   +  \epsilon_W~.
\label{eq:power1}
\ee
The effective chiral dimension of  vertex $i$ is given by  
$\epsilon_i = d_i  + n_i/2  - 2 \geq 0$,  where  $d_i$ is  the chiral dimension of the vertex (e.g.  a vertex  from ${\cal L}_{MB}^{(1)}$ has 
$d_i = 1$),  and $n_i$ is the number of baryonic legs attached to the vertex. 
Note that in \eq{eq:power1} we have explicitly isolated the contribution $\epsilon_W$  due to the weak vertex involving the 
external source coupled to the WIMP. 
In Table~\ref{tb1} we give a summary of chiral dimensions for the relevant  effective Lagrangians. 

\begin{figure}[!t]
\center\includegraphics[width=0.8\textwidth]{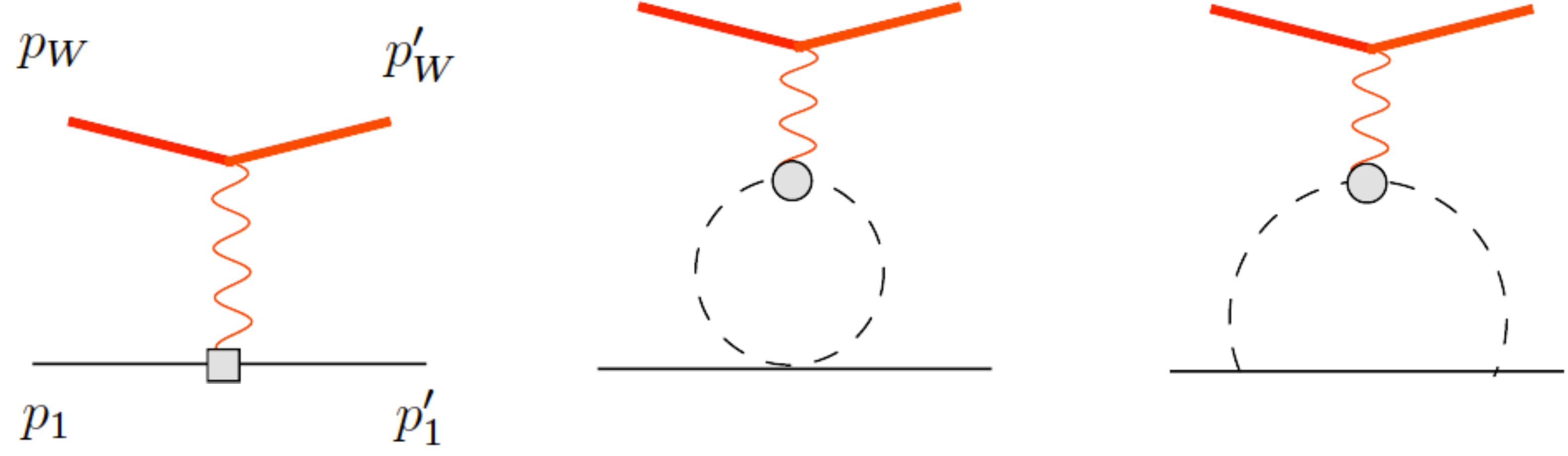}
\caption{LO and NLO diagrams contributing $M_{1,W}$. Black solid (dashed)  lines denote nucleons (mesons).}
\label{fig:diagrams}
\end{figure}

For fixed $A$, the leading contributions  to the amplitude are obtained by minimizing $\nu$ in \eq{eq:power1}, 
which is obtained by:
(i)  maximizing the number of connected contributions $C = A, A-1,  \dots$; 
(ii)  minimizing the number of loops $L=0,1, \dots $; 
(iii) using strong vertices from the lowest order Lagrangians 
$\mathcal{L}_{M}^{(2)}$ and $\mathcal{L}_{MB}^{(1)}$ $(\epsilon_i=0$), 
so as to minimize the $\epsilon_i$;  
(iv)  attaching the external scalar source to a baryon line using a vertex 
from ${\cal L}_{MB}^{(2)}$   ($\epsilon_W = 1$)  
or to a meson line using a vertex from ${\cal L}_M^{(2)}$ ($\epsilon_W=0$), 
consistently with the requirement that there are no external meson lines 
and the choice of $C$ and $L$.   
In the case of external scalar source we find: 
\begin{itemize}
\item  The leading order diagrams have $C=A$,  $L=0$, and
$\epsilon_W = 1$,\footnote{Note that $\epsilon_W=0$ is not consistent with the choice $C=A$ and $L=0$, 
because the source would have to couple to a meson line and the meson line has to attach to nucleons. This 
produces either $L=1$ or $C=A-1$.} 
i.e.   they have  $A$ disconnected parts,   no loops, (i.e.  no mesons in the diagram),   and the source attached to one of 
the nucleon lines through the  vertex in ${\cal L}_{MB}^{(2)}$. 
This corresponds to   $\nu_{\rm LO} = 5 - 3 A$. 
\item Three classes of diagrams can contribute to NLO ($\nu = \nu_{\rm LO} + 1$) 
as can be seen by inspecting  \eq{eq:power1}:
(i) $C=A$, $L=1$,  $\epsilon_W=0$,  i.e. diagrams with $A$ disconnected parts, one of which 
involves a  one-loop diagram with vertices from  ${\cal L}_{MB}^{(1)}$ and 
the source attached to a  meson line through ${\cal L}_M^{(2)}$ (see Fig.~\ref{fig:diagrams}); 
(ii) $C=A-1$, $L=0$,  $\epsilon_W=0$, i.e. diagrams with  $A-1$ disconnected parts, one of which involves two nucleon lines 
connected by  meson exchange with vertices from  ${\cal L}_{MB}^{(1)}$  and source  attached to the meson line  
through ${\cal L}_M^{(2)}$  (see Fig.~\ref{fig:diagrams2});
(iii) $C=A$, $L=0$,  $\epsilon_W=2$,  i.e. same topology as the leading diagram but with 
 the source attached to a nucleon line through a vertex from the  $O(p^3)$ Lagrangian ${\cal L}_{MB}^{(3)}$. 
By inspecting  of the only relevant vertex~\cite{Bernard:1995dp} 
${\cal L}_{MB}^{(3)} \supset  {\rm  Tr} (\chi_+)  \, {\rm Tr} ( \bar{B} v \cdot \partial B)$, 
one sees that for on-shell nucleons this contribution actually scales as of $p^{ \nu_{\rm LO} + 2}$ and therefore enters at NNLO. 
\end{itemize}
The number of diagrams  grows quickly as one goes beyond NLO, and  higher order terms  will involve in general vertices 
from effective Lagrangians containing more than two baryon fields, not considered in Table~\ref{tb1}.  
In this work we consider only NLO contributions and note that a consistent chiral counting to NLO 
requires to include not only loop corrections to the nucleon scalar form factors  (Fig.~\ref{fig:diagrams}), but also 
meson-exchange diagrams that result in two-nucleon operators (Fig.~\ref{fig:diagrams2})

Finally, a  similar analysis can be done for the insertion of the energy-momentum tensor vertices, 
coupled to the external source $s_\Theta(x)$ (see \eq{eq:lag2}).
Using the observation that  insertions of the energy-momentum tensor on a  baryon line scale as 
$\Theta^{\mu}_{\mu} \sim O(p^{0}, p, \dots)$~\cite{Cirigliano:2012xx}  (corresponding to $\epsilon_W = -1,0,\dots$)
and  on a meson line as  $\Theta^{\mu}_{\mu} \sim O(p^{2}, p^4, \dots)$~\cite{Donoghue:1991qv}
(corresponding to $\epsilon_W = 0,2, \dots$), we find that the first chiral corrections to the relation
$\langle N | \Theta^\mu_\mu | N \rangle  =  m_N \bar{\psi}_N  \psi_N$  arise in principle  at NNLO. 
Moreover, an explicit calculation~\cite{Cirigliano:2012xx}  shows that the relevant diagrams cancel  to this order, 
thus pushing the corrections to N$^3$LO.

\section{NLO corrections}
\label{sect:nlo}

We now discuss  the NLO contributions to $M_{A,W}$, the $A$-nucleon irreducible amplitude in presence of 
one insertion of the external source.  As discussed earlier, the NLO corrections fall into two classes: loop diagrams contributing to single-nucleon
amplitudes and tree-level diagrams contributing to two-nucleon interactions.

\subsection{One-nucleon amplitude}\label{sec:loops}

The one-nucleon amplitude starts at  leading order (tree-level diagram with vertex from ${\cal L}_{MB}^{(2)} $ in \eq{eq:LMB2} above) 
and receives NLO corrections through one-loop diagrams, as shown in  Figure \ref{fig:diagrams}. 
Including NLO corrections,  and denoting  the HBChPT spinors for  proton and neutron with   $N^T  = (H_v^{(p)},  H_v^{(n)})$, 
the one-nucleon amplitude reads 
\be
M_{1,W} =   \bar{\chi} \chi \ \left[ \frac{1}{2} \Big(f_p (q^2)  + f_n (q^2)\Big)    \  \bar{N} N \ + \ 
\frac{1}{2} \Big(f_p (q^2)  - f_n (q^2)\Big)   \  \bar{N} \tau_3  N  \right]~,
\label{eq:M1W}
\ee
with proton and neutron form factors given by 
\be
f_{p/n}(q^2) = 
\frac{1}{v \, \Lambda^2_{\text{np}}} \, \left[ 
\ \sum_{q=u,d,s}  \  \lambda_q  \, \sigma_q^{(p/n)}  \ + \ \lambda_\Theta \, m_{p/n}   
\ - \ \frac{g_A^2}{64\pi \, F_{\pi}^2}\,  \Big( A(q^2)   \pm B (q^2)  \Big)
\ \right]~, 
\label{Ab}
\ee
where 
$A (0) = B(0) = 0$. 
$f_{p/n} (0)$ receive contributions to LO   
(in terms of the  couplings  $b_0, b_D, b_F$  appearing  in \eq{eq:LMB2})
and NLO (loops) in the chiral expansion. 
Here we have chosen to lump  these contributions  in the sigma-terms 
defined by  $\bra{i}m_qq\bar{q}\ket{i} = \sigma_q^{(i)}  \bar{\psi}_i \psi_i$. 

The $\sigma_q^{(i)}$  ($i=p,n$)  can be   
expressed in terms of the  matrix elements 
$\sigma_{\pi N} =( (m_u + m_d)/2  )  \langle p | \bar{u} u + \bar{d} d |p \rangle$, 
$\xi  =   \langle p | \bar{u} u -  \bar{d} d |p \rangle/  \langle p | \bar{u} u + \bar{d} d |p \rangle$,  
$y = 2 \langle p |\bar{s} s | p\rangle /  \langle p | \bar{u} u + \bar{d} d |p \rangle$, 
and ratios of the light quark masses (see for example \cite{Corsetti:2000yq}).  
$\sigma_{\pi N}$ and $y$ can be extracted phenomenologically from baryon masses  and meson-baryon 
scattering data~\cite{Borasoy:1996bx,Pavan:2001wz}  or can be 
computed within lattice QCD~\cite{Ohki:2008ff,Young:2009zb,Toussaint:2009pz,Takeda:2010cw,Durr:2011mp,Horsley:2011wr}
(see Ref.~\cite{Kronfeld:2012uk} for a recent review), 
while $\xi$ can be related to  $y$ through an analysis of  baryon masses in the  $SU(3)$ limit~\cite{Cheng:1988im}.
In our analysis  we use the same relations  of  Ref.~\cite{Corsetti:2000yq}, 
but with updated numerical input on $\sigma_{\pi N}$,   $\sigma_s^{(p)}$   
(for which we use the ranges 
$\sigma_{\pi N}=  (45 \pm 15)$~MeV  and $\sigma_s^{(p)}=(45 \pm 25)$~MeV~\cite{Kronfeld:2012uk}) 
and the ratios of quark masses  (for which we use the PDG values~\cite{Nakamura:2010zzi}).

The momentum-dependent part of the  form factors arise to this order entirely from the one-loop diagrams, 
and depend on the lowest order couplings $F,D$ of ${\cal L}_{MB}^{(1)}$  in \eq{eq:LMB1} 
through the combinations $g_A = D+F = 1.27$~\cite{Nakamura:2010zzi} and
$\alpha= F/(D+F) \approx 0.4$~\cite{Borasoy:1996bx} (for which we will use the range $\alpha \in [0.3,0.5]$). 
Using the on-shell condition for external heavy baryons,  the  diagrams in Figure \ref{fig:diagrams} are  finite. 
Defining  ${\lambda}_{\pm} \equiv (m_u \lambda_u\pm m_d\lambda_d)/(m_u+m_d)$ and   $x_M = - q^2/m_M^2$,  the form factors are given by:
\footnote{We use the leading-order 
mass relations  with $m_u=m_d$  
to express the products $B_0 m_q$ in terms of  meson masses  
$m_{\pi,K,\eta}$.  On the other hand,   we keep $m_u \neq m_d$ in the overall factors $\lambda_{\pm}$. 
This prescription allows us to keep terms of order $(m_u - m_d)/(m_u + m_d) \sim O(1)$, 
while neglecting terms  of order $(m_u - m_d)/m_s \ll1$. }
\begin{eqnarray}
A(q^2) &=& \ 3 \, m_{\pi}^3\,{\lambda_{+}} \, \bar{f}(x_{\pi})
+
\Bigg(\frac{m_{\pi}^2}{3}\, \lambda_{+}\  +\ \frac{4\left(m_K^2-\frac{1}{2}m_{\pi}^2\right)}{3}{\lambda_{s}}\Bigg)\left(\frac{1-4\alpha}{\sqrt{3}}\right)^2 m_{\eta}\ \bar{f}(x_{\eta})
\nonumber\\
&+& \  \left(\frac{m_{\pi}^2}{2}   \, \lambda_{+} +\left(m_K^2-\frac{1}{2}m_{\pi}^2\right){\lambda_s}\right)\left(3(1-2\alpha)^2+\left(\frac{1+2\alpha}{\sqrt{3}}\right)^2\right)m_K \, \bar{f} (x_K)
\label{Adef}
\\
B(q^2) &=& 
\ 
m_\pi^2 \,\lambda_{-}  \ 
\Bigg[
\frac{1}{2}\left(\left(\frac{2\alpha+1}{\sqrt{3}}\right)^2-(1-2\alpha)^2\right)m_K  \, \bar{f}(x_K) 
\nonumber \\
&-&\ \frac{1-4\alpha}{3} \, (m_{\eta}+m_{\pi})  \, \bar{f}_1(x_{\eta}, x_{\pi})
\Bigg].
\label{Bdef}
\end{eqnarray}
The loop functions describing  the running of mesons with equal or different masses inside the loops are given by 
$\bar{f}(x) = f(x) - f(0)$ and $\bar{f}_1 (x_1,x_2) = f_1(x_1,x_2) - 4 (1 + z +z^2)/(1 +z)^2$ ($z= m_\pi/m_\eta$), respectively, with 
\begin{eqnarray}
f\left(x\right) & =& 2+\frac{(2+x)}{\sqrt{x}}\,\text{ArcCot}\left(\frac{2}{\sqrt{x}}\right),\label{eq:fdef}\\
f_1\left(x_1,x_2\right)&=&2+\frac{\left(x_1+x_2+x_1 x_2\right)\Bigg(\text{ArcCot}\left[\frac{2\sqrt{x_1}x_2}{x_1+x_1 x_2-x_2}\right]+\text{ArcCot}\left[\frac{2\sqrt{x_2}x_1}{x_2+x_1 x_2-x_1}\right]\Bigg)}{x_1\sqrt{x_2}+\sqrt{x_1}x_2}~, 
\end{eqnarray}
and  have the following useful properties:  $f(x\rightarrow 0)=3+\frac{5 x}{12}+\mathcal{O}(x^2)$, 
$f_1(x,x)\equiv f(x)$.

\begin{figure}[!t]
\center\includegraphics[width=0.5\textwidth]{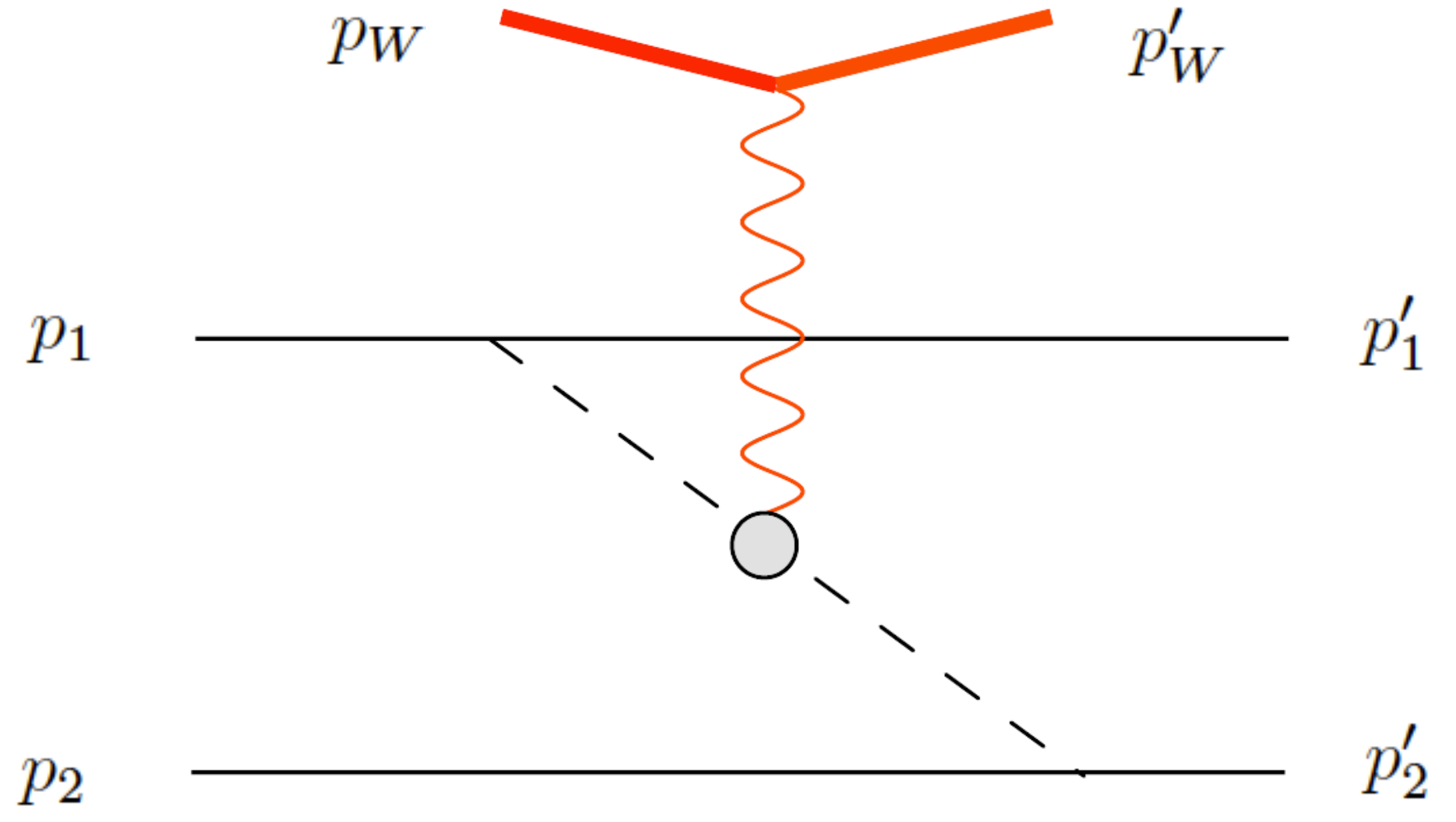}
\caption{Tree-level diagram contributing to $M_{2,W}$. Black solid (dashed)  lines denote nucleons (mesons).}
\label{fig:diagrams2}
\end{figure}

\subsection{Two-nucleon amplitude}\label{sec:TwoBodyOperators}

As was derived in Section \ref{sec:chpt}, at NLO 
there appears  also a contribution with  $A-1$ tree-level disconnected nucleon sectors, one of which involves 
two nucleons and the external source. The relevant diagram is shown in Figure \ref{fig:diagrams2}, and 
the possible mesons that are exchanged are limited to $\pi$  and $\eta$.
The corresponding  ``direct''  connected amplitude reads ($q_i = p_i - p_i'$ denotes the four-momentum transfer 
for each nucleon): 
\begin{eqnarray}
M_{2,W} &=&\mathcal{M}_{\pi\pi}+\mathcal{M}_{\eta\eta},
\label{eq:M2W}
\\
\mathcal{M}_{\pi\pi}&=&
-\frac{1}{v\,\Lambda_{\text{np}}^2}  \, \frac{g_A^2}{F_\pi^2}\,
\,\frac{m_{\pi}^2\,\lambda_{+}}{(q_1^2-m_{\pi}^2)(q_2^2-m_{\pi}^2)}  \  
\bar{N}   q_1 \cdot S   \, \vc{\tau}_1^k N \  \bar{N} q_2 \cdot S   \,  \vc{\tau}_2^k N    \  \bar{\chi} \chi~, 
\label{eq:pipi}\\
\mathcal{M}_{\eta\eta}&=&
-\frac{1}{v\,\Lambda_{\text{np}}^2}  \, \frac{g_A^2}{3 \, F_\pi^2}   \left(\frac{4\alpha-1}{\sqrt{3}}\right)^2  
\frac{m_{\pi}^2  \,\lambda_{+}+4\left(M_K^2-\frac{1}{2}m_{\pi}^2\right)\lambda_{s}}{(q_1^2-m_{\eta}^2)(q_2^2-m_{\eta}^2)}  
\  \bar{N}  q_1 \cdot S N  \  \bar{N}  q_2 \cdot S N \  \bar{\chi} \chi~.  \qquad
\label{eq:etaeta}
\end{eqnarray}
There is also an ``exchange'' amplitude, which is obtained from the direct one by  
changing the overall sign and interchanging  
all variables of the final-state nucleons ($p_1'  \leftrightarrow p_2'$,  $N_1'  \leftrightarrow N_2'$). 

Compared to Ref. \cite{Prezeau:2003sv}, where the $\pi\pi$ two-body interaction has been calculated, 
we  also include the $\eta \eta$  which is  Yukawa  enhanced as can be seen from \eq{eq:etaeta}. 
Later on we will  study the competition between this enhancement and the suppression 
expected from the fact that the $\eta$-induced potential has shorter range compared to the $\pi$-mediated one.

\subsection{WIMP-nucleon potentials}

From the WIMP-nucleon amplitudes discussed above to NLO in the chiral power counting, 
one can derive non-relativistic WIMP-nucleon interaction potentials. 
This procedure  is standard and requires (i)  taking the non-relativistic limit of $M_{1,W}$ and $M_{2,W}$ 
and rescaling  the amplitude to take into account the difference between relativistic and non-relativistic 
normalization of states:  $M_{A,W} \to  \overline{M}_{A,W} \equiv 1/ ((2 m_N)^A  \, (2 m_W)) \,  M_{A,W} |_{\rm non-rel}$~; 
(ii) determining  the potential by requiring that the  amplitude of non-relativistic quantum mechanics 
matches $\overline{M}_{A,W}$,  using the same (unbound) WIMP and nucleon external  states used in the 
ChPT calculation. This step involves taking the Fourier transform of  $\overline{M}_{A,W} (\vec{q}_i, \vec{q}_W)$. Note that
for the two body term one can work  exclusively with the ``direct'' amplitude and match it on the 
QM matrix element of the potential between nucleon wave-functions that are not anti-symmetrized. 
The same result would be obtained by including the
``exchange" diagram and using an anti-symmetric two-body wave-function.

The matching procedure leads to a QM interaction Hamiltonian of the form: 
\be 
H_I \  = \ \sum_{i=1,A} \ V_1 (\vec{x}_i - \vec{x}_W)  \ + \  \sum_{i < j}   \ V_2 (\vec{x}_i - \vec{x}_W,  \vec{x}_j - \vec{x}_W)  \ + \ \dots ~, 
\label{eq:HWN}
\ee
where the 1-body ($V_1$) and 2-body  ($V_2$) terms  are related to $\overline{M}_{1,A}$  and $\overline{M}_{2,A}$, respectively.
The LO matching induces  a contact one-body potential $V_1 (\vec{x}_i - \vec{x}_W) \propto  \delta^{(3)}  (\vec{x}_i - \vec{x}_W)$. 
The NLO matching induces a long-range term in the WIMP-nucleon  
potential $V_1 (\vec{x}_i - \vec{x}_W)$,  as well as a contribution to the WIMP-nucleon-nucleon potential. 
As discussed below, we shall not need the detailed expression of the coordinate-space potentials, but 
rather a hybrid form that depends on the nucleon coordinates 
$\vec{x}_i$ (and spin-isospin variables)  and the  WIMP momentum-transfer variable $\vec{q}_W$.

\section{Nuclear matrix elements}  
\label{sec:NSM}

The next step towards a description of WIMP-nucleus elastic scattering requires using 
the WIMP-nucleon(s) interaction Hamiltonian (\ref{eq:HWN}) to compute the transition amplitude 
between appropriate initial and final WIMP-nucleus states. 
The initial state  $ | i \rangle$ is characterized by a WIMP  with momentum $\vec{P}_W$ (and possibly internal quantum numbers) 
and nucleus in its ground state  with center-of-mass   momentum $\vec{P}_A$.
The final state $|f\rangle$  is characterized by momenta $\vec{P}_W ' = \vec{P}_W - \vec{q}_W$  and $\vec{P}_A ' = \vec{P}_A - \vec{q}_A$, 
with the nucleus remaining in the ground state. 
Using translational invariance,  the  nucleus wave-function in the initial state can be written as 
$\psi_i (\vec{x}_1, \dots, \vec{x}_A) =  e^{i \vec{P}_A \cdot \vec{X}_{CM}}  \phi_0 (\vec{y}_1, \dots, \vec{y}_{A-1})$, 
where  for simplicity we have suppressed  spin and isospin indices, 
$\vec{X}_{CM}, \vec{y}_1, \dots, \vec{y}_{A-1}$ represent center-of-mass and internal (Jacobi) coordinates 
for the system of $A$ nucleons, and $\phi_0  (\vec{y}_1, \dots, \vec{y}_{A-1})$ is the ground-state wave-function in 
internal coordinates.

Within this  setup, 
one can show~\cite{Cirigliano:2012xx} that the non-relativistic $T$-matrix element reads
\begin{eqnarray} 
\langle f | \hat{T} | i \rangle &=&  (2 \pi)^3  \delta^{(3)} \,  (\vec{q}_W + \vec{q}_A) \ T (\vec{q}_W)  \qquad \qquad  \qquad 
 T (\vec{q}_W)  = T_{1} + T_{2}  
\\
T_1  &=&
 \sum_{i=1,A}  \ \int \, d\vec{x}_1 \dots d\vec{x}_A \  \ 
\psi_0^* (\vec{x}_1, \dots, \vec{x}_A) \, \otimes  \, \tilde{V}_1 (\vec{q}_W;  \vec{x}_i) 
\, \otimes  \,   \psi_0 (\vec{x}_1, \dots, \vec{x}_A) 
\qquad
\label{eq:T1}
\\
T_2 &=& 
 \sum_{i<j}  \ \int \, d\vec{x}_1 \dots d\vec{x}_A \  \ 
\psi_0^* (\vec{x}_1, \dots, \vec{x}_A) \,  \otimes  \,   \tilde{V}_2 (\vec{q}_W;  \vec{x}_i, \vec{x}_j) 
\, \otimes  \,   \psi_0 (\vec{x}_1, \dots, \vec{x}_A) ~, 
\qquad 
\label{eq:T2}
\end{eqnarray}
where we indicate with the symbol ``$\otimes$'' the non-trivial contractions in spin and isospin space. 
In the above expressions 
$ \psi_0 (\vec{x}_1, \dots, \vec{x}_A)$ denotes the ground-state nuclear wave-function with 
center-of-mass localized at the origin, 
i.e.   $\psi_0 (\vec{x}_1, \dots, \vec{x}_A) =  \psi_{CM} (\vec{X}_{CM})  \, \phi_0 (\vec{y}_1, \dots, \vec{y}_{A-1})$,
with $|\psi_{CM} (\vec{X}_{CM})|^2 \propto  \delta^{(3)} (\vec{X}_{CM})$.  
The hybrid potentials $\tilde{V}_{1,2}$  are related to the  one- and two-body amplitudes 
(obtained by taking the non-relativistic limit of  (\ref{eq:M1W}) and (\ref{eq:M2W})) 
as follows:
\begin{eqnarray}
\tilde{V}_1 (\vec{q}_W;  \vec{x}_i) &=& - e^{i \vec{q}_W \cdot \vec{x}_i}  \ \overline{M}_{1,W} (\vec{q}_W) 
\label{eq:Vt1}
\\
\tilde{V}_2 (\vec{q}_W;  \vec{x}_i, \vec{x}_j) &=& - 
\int \frac{d\vec{q}_i}{(2\pi)^3} \frac{d\vec{q}_j}{(2\pi)^3}  e^{- i \vec{q}_i \cdot \vec{x}_i}  e^{- i \vec{q}_j \cdot \vec{x}_j}  \ 
(2 \pi)^3  \delta^{(3)}  (\vec{q}_i + \vec{q}_j + \vec{q}_W) \  \overline{M}_{2,W} (\vec{q}_i, \vec{q}_j, \vec{q}_W) ~. 
\label{eq:Vt2}
\end{eqnarray} 
The explicit form of the two-body potentials at $\vec{q}_W=0$ is given by:
\begin{eqnarray}
\tilde{V}_2^{(\pi \pi)} (0 ;  \vec{x}_i, \vec{x}_j) &=&   
- \frac{\lambda_{+}}{v\,\Lambda_{\text{np}}^2}  \, \frac{g_A^2 m_\pi^3}{96  \, \pi \,  F_\pi^2}\, O_{\pi \pi}   (i,j)
\\
\tilde{V}_2^{(\eta \eta)} (0 ;  \vec{x}_i, \vec{x}_j) &=&   
- \frac{1}{v\,\Lambda_{\text{np}}^2}  \, \frac{g_A^2 m_\eta}{ 288 \, \pi \,  F_\pi^2}\ 
  \left(\frac{4\alpha-1}{\sqrt{3}}\right)^2   
\left[ m_{\pi}^2  \,\lambda_{+}+4\left(M_K^2-\frac{1}{2}m_{\pi}^2\right)\lambda_{s} \right]
\,  O_{\eta \eta}   (i,j)
\qquad
\end{eqnarray} 
with 
\begin{eqnarray}
{O}_{\pi\pi} (i,j) &=&-\frac{1}{x_{\pi}}\left(F_1(x_{\pi})\delta_{ab}+F_2(x_{\pi})T_{ab}\right) \left({\vc{\sigma}}_i^a\vc{\sigma}^b_j\right)\otimes\left(\vc{\tau}_{i}\mcdot
\vc{\tau}_j\right) ,\label{eq:pipiQM2}\\
{O}_{\eta\eta} (i,j)&=&-\frac{1}{x_{\eta}}\left(F_1(x_{\eta})\delta_{ab}+F_2(x_{\eta})T_{ab}\right) 
\left({\vc{\sigma}}_i^a\vc{\sigma}^b_j\right)\otimes\left(\mathbb{I}\right)   ,\label{eq:etaetaQM2}
\end{eqnarray}
and the definitions 
$ \vc{ \rho} = \vc{x}_i - \vc{x}_j$,  $T_{ij} = 3 \hat{\vc{\rho}}_i \hat{\vc{\rho}}_j -  \delta_{ij}$, 
$x_{\pi}=m_{\pi}|\vc{\rho}| ,x_{\eta}=m_{\eta}|\vc{\rho}|$, $F_1(x)=\e^{-x}(x-2), F_2(x)=\e^{-x}(x+1)$~\cite{Vergados:1981bm,  Prezeau:2003sv}.

Evaluating the above matrix elements requires  knowledge of  the nuclear many body wave function $| \psi_0 \rangle$. 
In particular,  evaluating the one-body  (\ref{eq:T1})  and two-body contributions   (\ref{eq:T2}) 
requires knowledge of the one-  and two-body nucleon  densities in the ground state.  
For typical nuclei involved in DM direct detection experiments such as Ge or Xe the relevant 
wave function cannot be obtained from first principles and different models have to be used.

The one-body contributions  to the WIMP-nucleus amplitude can be evaluated in 
a straightforward way noting  that,  given  the one-body potential from  (\ref{eq:Vt1}), 
the matrix element  (\ref{eq:T1}) factorizes in the product of  nucleon scalar  form factors 
and nuclear form factor (the Fourier transform of the one-body nucleon densities). 
Denoting the latter by  $F_{n,p}  (|\vec{q}_W |^2)$ and assuming that neutron and proton 
densities are equal, $F_{n,p}  (|\vec{q}_W|^2) \equiv F  (|\vec{q}_W|^2)$,  one finds:
\be
T_1   = -  F   (|\vec{q}_W|^2) \,  \Big( Z  \, f_p  (|\vec{q}_W|^2)  \ + \  (A-Z)  \, f_n  (|\vec{q}_W|^2)  \Big)~,
\label{eq:T1v2}
\ee
with $f_{p,n}$ given in \eq{Ab}. The  one-body density can be taken from phenomenology or from 
microscopic models, such as the nuclear shell model.  
We will use the  exponential form~\cite{Jungman:1995df} 
$F (E_R) = {\rm Exp} (- E_R /(2 E_0))$ with  $E_R = | \vec{q}_W|^2/(2 m_A)$, 
$E_0 = 1.5/(m_A  R_0^2)$ and $R_0 = [0.3 + 0.91 (m_A/{\rm GeV})^{1/3}] \times 10^{-13}$~cm. 
We have checked that the results are stable if we use 
other parameterizations available in the literature~\cite{Engel:1992bf,Jungman:1995df}.

In order to calculate the two-body contribution  $T_2$ (\ref{eq:T2})  to the WIMP-nucleus amplitude, 
one needs the matrix elements
\be
   {\cal N}_{MM} =  \langle \psi_0 | \ \sum_{i<j}  O_{MM}  (i,j) \  | \psi_0 \rangle  \qquad \qquad M = \pi, \eta 
\ee
of the  operators ${O}_{\pi\pi}$ and ${O}_{\eta\eta}$ defined in \eq{eq:pipiQM2} and \eq{eq:etaetaQM2}, 
in terms of which one has
\begin{eqnarray}
T_2^{(\pi \pi)} &=& - \frac{\lambda_{+}}{v\,\Lambda_{\text{np}}^2}  \, \frac{g_A^2 m_\pi^3}{96  \, \pi \,  F_\pi^2} \ {\cal N}_{\pi \pi} 
\label{eq:T2pi}
\\
T_2^{(\eta \eta)} &=& 
- \frac{1}{v\,\Lambda_{\text{np}}^2}  \, \frac{g_A^2 m_\eta}{ 288 \, \pi \,  F_\pi^2}\ 
  \left(\frac{4\alpha-1}{\sqrt{3}}\right)^2   
\left[ m_{\pi}^2  \,\lambda_{+}+4\left(M_K^2-\frac{1}{2}m_{\pi}^2\right)\lambda_{s} \right]
\,  {\cal N}_{\eta \eta}  ~.  
\label{eq:T2eta}
\end{eqnarray}
To evaluate the matrix elements   $\mathcal{N}_{\pi\pi, \eta \eta}$  we use  the NSM. 
In this framework,  one  assumes that the nucleons feel 
a mean external potential  and occupy 
levels according to Pauli's exclusion principle.  
For the self-consistent potential, we use 
the  harmonic oscillator with nucleus-dependent 
frequency $\omega(A)$  empirically fit to data~\footnote{We use  the following form for the harmonic oscillator frequency: $\omega(A)=(45/A^{1/3}-25/A^{2/3})\text{MeV}$ \cite{AnnaHayes2012}. 
For the lower cutoff in the radial integrals we use $0.5$~fm.}.
Given an arbitrary two-body potential $V_{ij}$ between nucleons $i$ and $j$, 
NSM allows to calculate the expectation value of the following Hamiltonian: $G=\sum_{i<j}V_{ij}$. For the simplest case of all closed shells (core-core matrix element), using the raising and lowering operator formalism the result for such expectation value in NSM equals:
\begin{eqnarray}
\bra{c}G\ket{c}=\sum_{j1\le j_2, J, T}(2J+1)(2T+1) V_{JT}(j_1, j_2, j_1, j_2),
\end{eqnarray}
where $j_i$ represent the orbits of NSM, encoding quantum numbers $n,l,j$ of the orbit, $j_1\le j_2$ is understood in the sense $E_{j_1}\le E_{j_2}$, and finally $J$ runs $|{j_1-j_2}|...j_1+j_2$, $T=0,1$. $V_{JT}$ is a two-body matrix element between the anti-symmetrized two-body wavefunction. Explicit expression for such two-body matrix element can be found in Ref. \cite{Castel}.
By computing the matrix elements for a number of closed-shell nuclei, we find the following scaling with $A$
\begin{eqnarray}
\mathcal{N}_{\pi\pi}\approx-0.91 A, \qquad\mathcal{N}_{\eta\eta}\approx 0.0061A ~.
\end{eqnarray}
For ${\cal N}_{\pi \pi}$ the scaling is consistent  with  $\mathcal{N}_{\pi\pi}\sim A$ found in \cite{Prezeau:2003sv}.
The sign difference between $\mathcal{N}_{\pi\pi}$ and $\mathcal{N}_{\eta\eta}$ appears because for the pion exchange diagram the second term  
in the scalar form-factor $F_1(x)=(x-2)\exp(-x)$ dominates, while for $\eta\eta$ on the contrary, the first term dictates the sign.
In order to understand the size difference one has to compare $m_{\pi}\,\mathcal{N}_{\pi\pi}$ to $m_{\eta}\,\mathcal{N}_{\eta\eta}$ since the meson mass $m_{M}$ has been factored out in the definitions of nuclear operators $\mathcal{O}_{MM}$. The ratio of these numbers approximately equals $37$. This arises from a factor of 3 suppression for $\eta\eta$ operator due to different 
isospin structure and a  factor of  $\approx  12$  due to shorter distance potential for $\eta\eta$ compared to $\pi\pi$ exchange. 
Thus the  expected value for the ratio $T_{2}^{(\eta\eta)}/T_{2}^{(\pi\pi)}$ is $1/37  \times 20 \times (4\alpha-1)^2/3  \approx 0.07$, 
where the factor of $\approx 20$ arises due to the strange Yukawa enhancement factor for the  $\eta\eta$  operator.

\section{Phenomenology}\label{sec:phenomenology}

The differential WIMP-nucleus scattering rate  per unit time and unit detector mass  reads 
\be
\frac{dR}{dE_R}=\frac{ \kappa_W  \rho_W}{\pi\,m_{W}} \,  \bigg| \Big[Z f_{p}(E_R)+(A-Z)f_n(E_R)\Big] F(E_R) - 
 T_2(E_R,A,Z) \bigg|^2\,\eta\left(E_R,m_W,m_A\right) ~,
\label{eq:rate1}
\ee
where $E_R$ denotes  the nuclear recoil energy, related to the  momentum transfer squared via 
$- q_W^2 \simeq |\vec{q}_W|^2 =  2  m_A  E_R$  ($m_A$ is the nucleus mass).  
The overall factor  $\kappa_W$  in (\ref{eq:rate1})  depends on the nature of the DM particle.  
For example, for Dirac fermions $\kappa_W = 1/2$, while for Majorana fermions $\kappa_W = 2$. 
We denote by $m_W$ the WIMP mass and $\rho_W$ the local dark matter density 
(for which we use $\rho_W=~0.3\,\text{GeV}/\text{cm}^3$).  
Next, the WIMP-nucleus scattering amplitude is given by  the sum of a  one-body term  
involving  the nucleon form factors   $f_{p,n} (E_R)$ (see \eq{Ab})
and nuclear form factor $F(E_R)$  (see discussion following (\ref{eq:T1v2})), 
and a two-body term $T_2 (E_R,A,Z)$ (see \eq{eq:T2pi} and \eq{eq:T2eta}). 
Finally,  the last factor in  (\ref{eq:rate1})  involves  an integral over the local DM velocity distribution $f(u)$: 
\be
\eta\left(E_R,m_W ,m_A\right) = \int_{u_{\text{min}}}^{u_{\text{esc}}}\frac{f (u)}{u}\text{d}^3\vc{u}~,  \qquad 
u_{\rm min} = \sqrt{\frac{m_A E_R}{2 \mu_{WA}^2}}
\qquad  
\mu_{WA} = \frac{m_W m_A}{m_W + m_A}~.
\ee
In our study  we use for illustrative purposes a Maxwellian distribution with  finite escape velocity 
(for which analytic expressions can be found in Ref.~\cite{Savage:2006qr}), with input parameters 
$v_0=  220\,\text{km}/\text{sec}$, 
$v_{\text{obs}} =  233\,\text{km}/\text{sec}$, 
$v_{\text{esc}} = 550\,\text{km}/\text{sec}$~\cite{Savage:2006qr}.

Fixing the hadronic  parameters to the central values of the ranges  discussed  in Section~\ref{sec:loops}, 
we can express the nucleon form factors and the two-body amplitude as linear combinations 
of the  short-distance parameters  $\lambda_{u.d.s,\Theta}$  
defined above in \eq{eq:sqvslambdaq}.
At zero momentum  transfer the form factors read~\footnote{The coefficients of $\lambda_{u,d}$ 
are proportional to $\sigma_{\pi N}$, while the coefficient of $\lambda_s$ is $\sigma_s$, 
so one can immediately  assess the impact of hadronic uncertainties.}  
\begin{eqnarray}
f_p(0) \equiv f_p &=&\frac{1}{v\,\Lambda^2_{\text{np}}}\times \Big(17.7 \,\lambda_u+24.5\, \lambda_d + 45.0\, \lambda_s  + 
 938.3\, \lambda_{\Theta} \Big)  \times {\text{MeV}}  ,
 \label{eq:LO1}\\
f_n(0) \equiv f_n &=&\frac{1}{v\,\Lambda^2_{\text{np}}}\times  \Big( 12.2 \,\lambda_u+35.5\, \lambda_d + 45.0\, \lambda_s  + 939.5\,\lambda_{\Theta}\Big)   \times {\text{MeV}} .    \label{eq:LO2}
\end{eqnarray}
The energy-dependence of the form factors arising from loop corrections is very well approximated by 
a linear form  for $E_R < 50$~keV:~\footnote{
The uncertainty in $\alpha = F/(F+D) \in [0.3,0.5]$ affects the coefficients of $\lambda_{u,d}$ at the 5\% level, 
and the coefficient of $\lambda_s$ at the 20\% level.
} 
\begin{eqnarray}
f_p(E_R) - f_p(0) 
&\simeq &\frac{1}{v\,\Lambda^2_{\text{np}}}\times\Big[\left( - 0.58\,\lambda_u-0.96\,\lambda_d - 
2.38\,\lambda_s \right)\text{MeV} \Big]\times 
\left( \frac{A}{100} \right) \left( \frac{E_R}{50 \, \text{KeV}}\right)   \ \ \  
\label{eq:loop1}\\
f_n(E_R) - f_n(0) 
&\simeq &\frac{1}{v\,\Lambda^2_{\text{np}}}\times\Big[\left( - 0.48\,\lambda_u-1.16\,\lambda_d - 
2.38\,\lambda_s \right)\text{MeV} \Big]\times 
\left( \frac{A}{100} \right) \left( \frac{E_R}{50 \, \text{KeV}}\right) .   \ \  \
\label{eq:loop2} 
\end{eqnarray}
Using the exact one-loop function or the above linear parameterization leads to 
differences in the rates at most of 0.1\% (the largest deviations 
occurs for Xenon target and $m_W > 100$~GeV). 
Finally, the two-body  amplitude  is given by
\begin{eqnarray}
-T_{2}(0, A, A/2)^{\text{NSM}}_{\text{closed shells}}
&=&\frac{1}{v\,\Lambda^2_{\text{np}}}\times A\times  \Big( - 0.48 \,\lambda_u-0.97\,\lambda_d + 0.089\,\lambda_s \Big) \times 
\text{MeV},\label{eq:deltanlo}
\end{eqnarray}
where we limited ourselves  to closed shells, $Z=A-Z=A/2$. 
While strictly speaking this formula is inapplicable for most target nuclei used in experiments, 
we will use it as a rough estimate of the two-body effect. 
An improved analysis should go beyond closed shells and include 
the dependence of the two-body amplitude on  $E_R$. 
We leave this to future work.

The  main novelty of our analysis stems from  including in \eq{eq:rate1}  the $E_R$-dependence of $f_{p,n} (E_R)$ and the two-body 
amplitude $T_2$, both of  which arise to NLO in the chiral power counting. 
These  long-distance QCD effects should be included in any consistent fit to direct DM detection searches because, as we show below, they can affect both the shape of the recoil spectrum and the total  rates in a non-trivial way. 
We wish to emphasize here a few  points:
\begin{itemize}
\item While the new effects are of the  natural size expected by  power counting (compare 
(\ref{eq:LO1})-(\ref{eq:LO2})  to  (\ref{eq:loop1}), (\ref{eq:loop2}), and (\ref{eq:deltanlo})),  
they become  very important 
in those regions of parameter space where  
the leading order contribution to the WIMP-nucleus amplitude  
are  suppressed (such as those realizing the  so-called  ``isospin-violating dark matter"  scenarios).

\item Note  that the  NLO corrections  depend on the recoil energy and have a different dependence on the short-distance  
parameters $\lambda_i$ than the LO contributions. As a result, 
the differential cross-section (\ref{eq:rate1}) does not factorize into a product of a cross-section 
$\sigma_p \times [ Z + (A - Z) f_n/f_p]^2$, depending only on short-distance parameters $\lambda_i$, and a term depending only on long-distance QCD and nuclear effects. Still, for the differential cross-section the astrophysical dependence in $\eta$ factorizes. The scattering rate, however, obtained from integrating (\ref{eq:rate1}) over an energy window, no longer exhibits a factorization into a product of
$\sigma_p \times [ Z + (A - Z) f_n/f_p]^2$  with a term that is schematically astrophysical ${\otimes}$ nuclear andÊ independent of the short-distance parameters.Ê 
Using factorization to compare positive and null results of different direct-detection experiments, independent of assumptions about the DM velocity distribution, may have to be re-examined.

\item  Perhaps  most importantly,  our results show that 
the scalar-mediated WIMP-nucleus  cross-section cannot be parameterized 
in terms of just two quantities, namely $f_{p}$ and $f_{n}$ or 
equivalently  the WIMP-proton cross-section  $\sigma_p \propto   m_p^2  f_p^2$ and the ratio  $r = f_n/f_p$.  
Starting from the short-distance interaction of \eq{eq:lag1}, 
the cross-section depends on  four parameters (in one-to-one correspondence 
with $\lambda_{u,d,s,\Theta}$). 
This calls for a more general analysis of data, that takes into account these additional degrees of  freedom. 
\end{itemize}

A convenient choice of independent parameters, that matches onto  the standard choice when neglecting 
NLO chiral corrections, is achieved as follows. 
First, we  observe that $f_{n,p}(E_R)$ and $T_2 (E_R,A,Z)$ are linear functions of $\lambda_{u,d,s,\Theta}/\Lambda_{\rm np}^2$ so that the 
rate is a homogeneous quadratic form in the $\lambda$'s.   
Next,  we can trade $\lambda_{u,d}$ for  $f_p$ and $r= f_n/f_p$, and finally we  can extract   $\lambda_\Theta$ as an overall factor. 
In conclusion, the four parameters controlling the rate are: 
(1)  $\lambda_\Theta/(v \Lambda_{\rm np}^2 )$, which sets the overall normalization; 
(2)  $f_p$, or equivalently\footnote{The convenience of this choice is apparent from equations \eq{eq:LO1} and \eq{eq:LO2}. It is also clear that $\bar{f}_p$ has dimensions of energy.}   $\bar{f}_p = v \Lambda_{\rm np}^2 \,   f_p/\lambda_\Theta$;   
(3) $r = f_n/f_p$,;  and  (4)  $\lambda_s/\lambda_\Theta$. 
The rate has the form $R \sim ( \lambda_\Theta/(v \Lambda^2_{\rm np}))^2   \times  Q(f_p, r f_p, \lambda_s/\lambda_\Theta)$, 
where $Q(x,y,z)$ is a quadratic form in $x,y,z$. 
Neglecting NLO corrections, only two independent parameters survive, namely $f_p$ 
(or equivalently $\sigma_p \propto   m_p^2  f_p^2$) and $r = f_n/f_p$, 
and the rate takes the simplified form $R \sim f_p^2 \,  [ Z + (A-Z) r]^2$.  
Note that any ratios of integrated rates  only depend on three parameters:  $\bar{f}_p$, $r$, and $\lambda_s/\lambda_\theta$, as 
the overall normalization cancels.

\begin{figure}[!t]
\begin{center}
\includegraphics[width=0.45\textwidth]{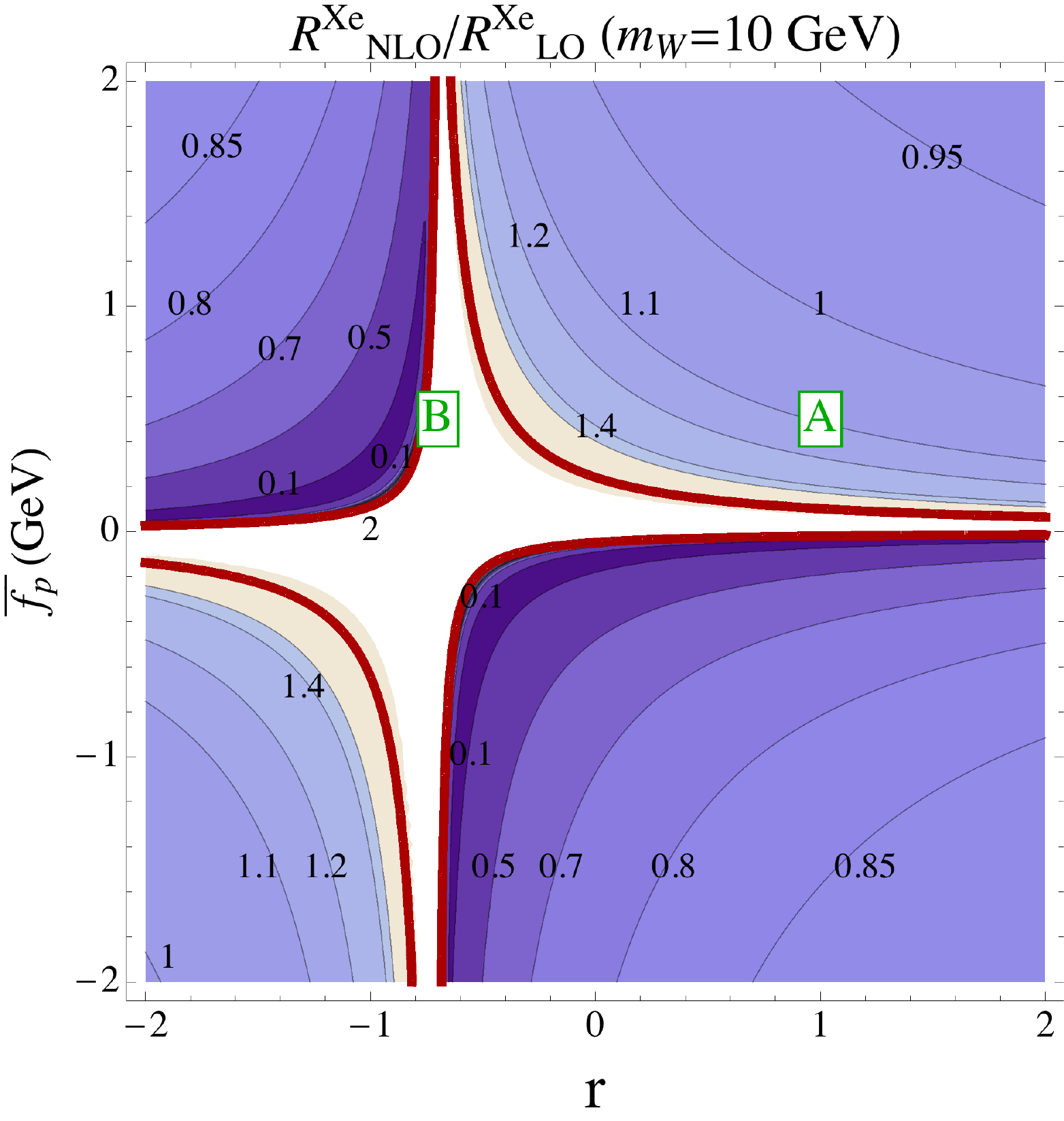} \hspace{0.05\textwidth} \includegraphics[width=0.45\textwidth]{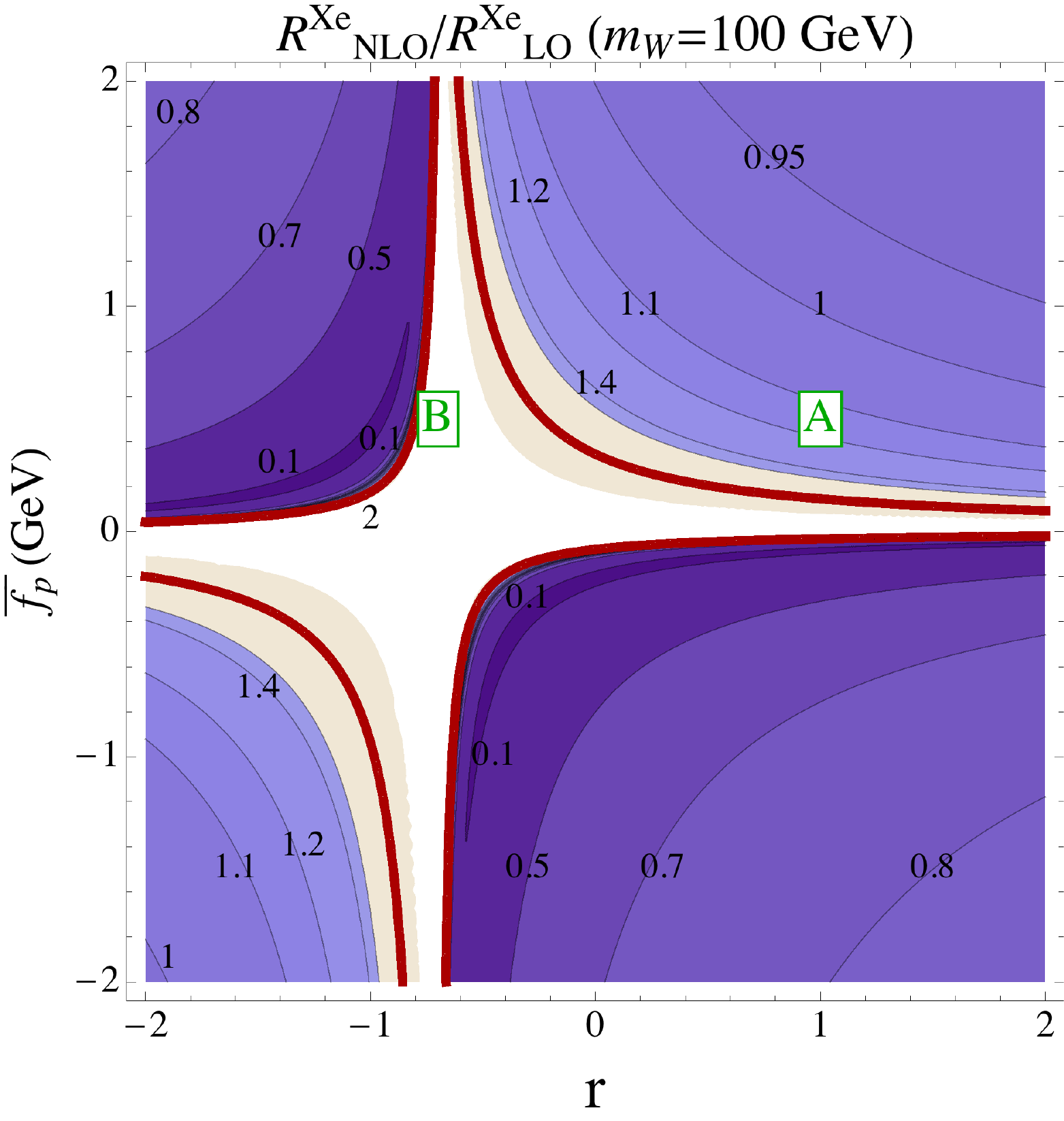}
\vspace{-.3cm}
\caption{\label{fig:pheno1} 
Contour plots of  the NLO to LO integrated rates  $R^{\rm Xe}_{\rm NLO}/R^{\rm Xe}_{\rm LO}$   on the  
$(r, \bar{f}_p)$ plane,   at fixed  $\lambda_s/\lambda_\Theta=1$, with  
$m_W = 10$~GeV  (left panel) and $m_W=100$~GeV (right panel). The solid red line corresponds to $R^{\text{Xe}}_{\text{NLO}}/R^{\text{Xe}}_{\text{LO}}=2$ and for all points inside the solid red line the NLO correction is more than 100\%. }
 \vspace{-0.5cm}
\end{center}
\end{figure}

\begin{figure}[!t]
\begin{center}
\includegraphics[width=0.37\textwidth]{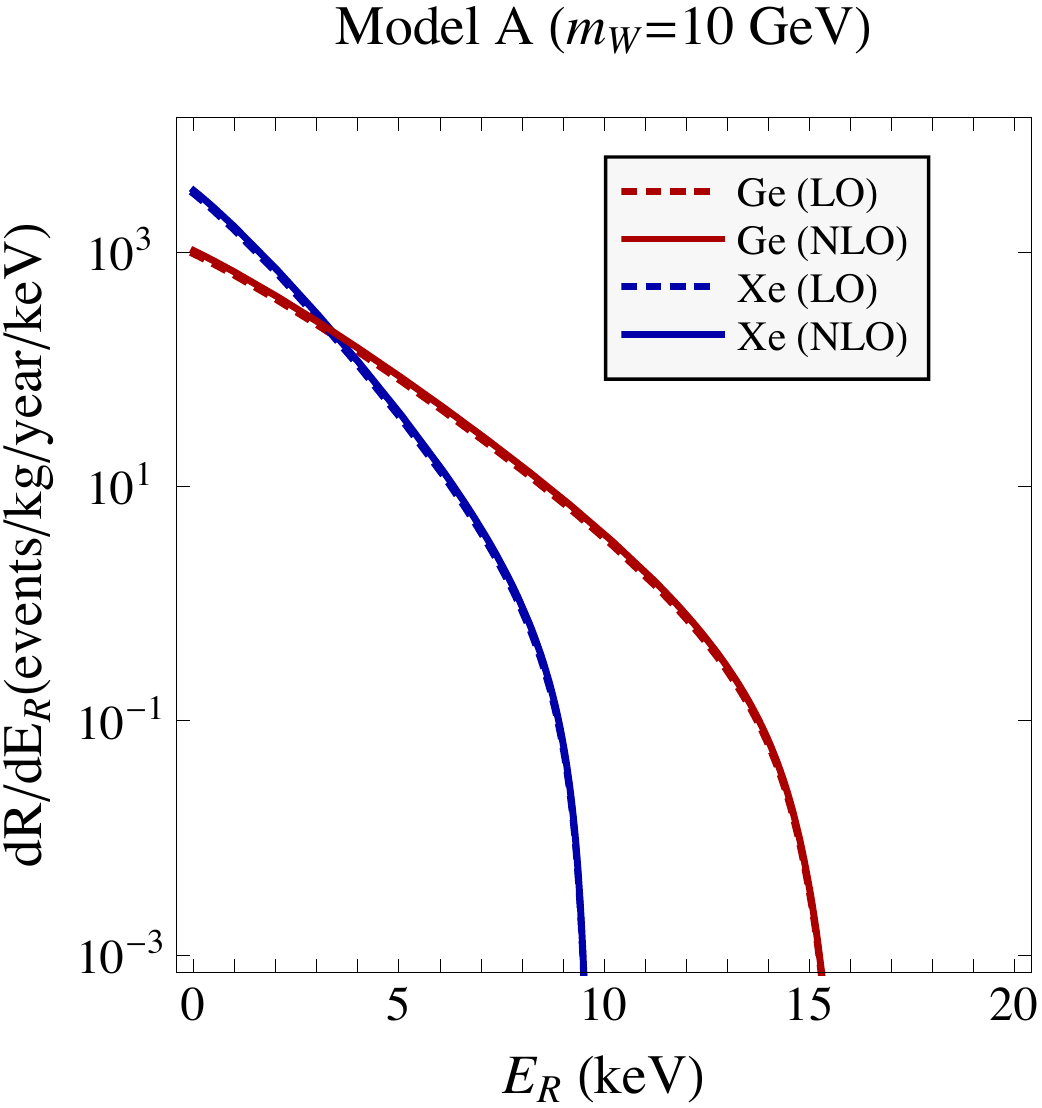}     \hspace{0.05\textwidth}    \includegraphics[width=0.362\textwidth]{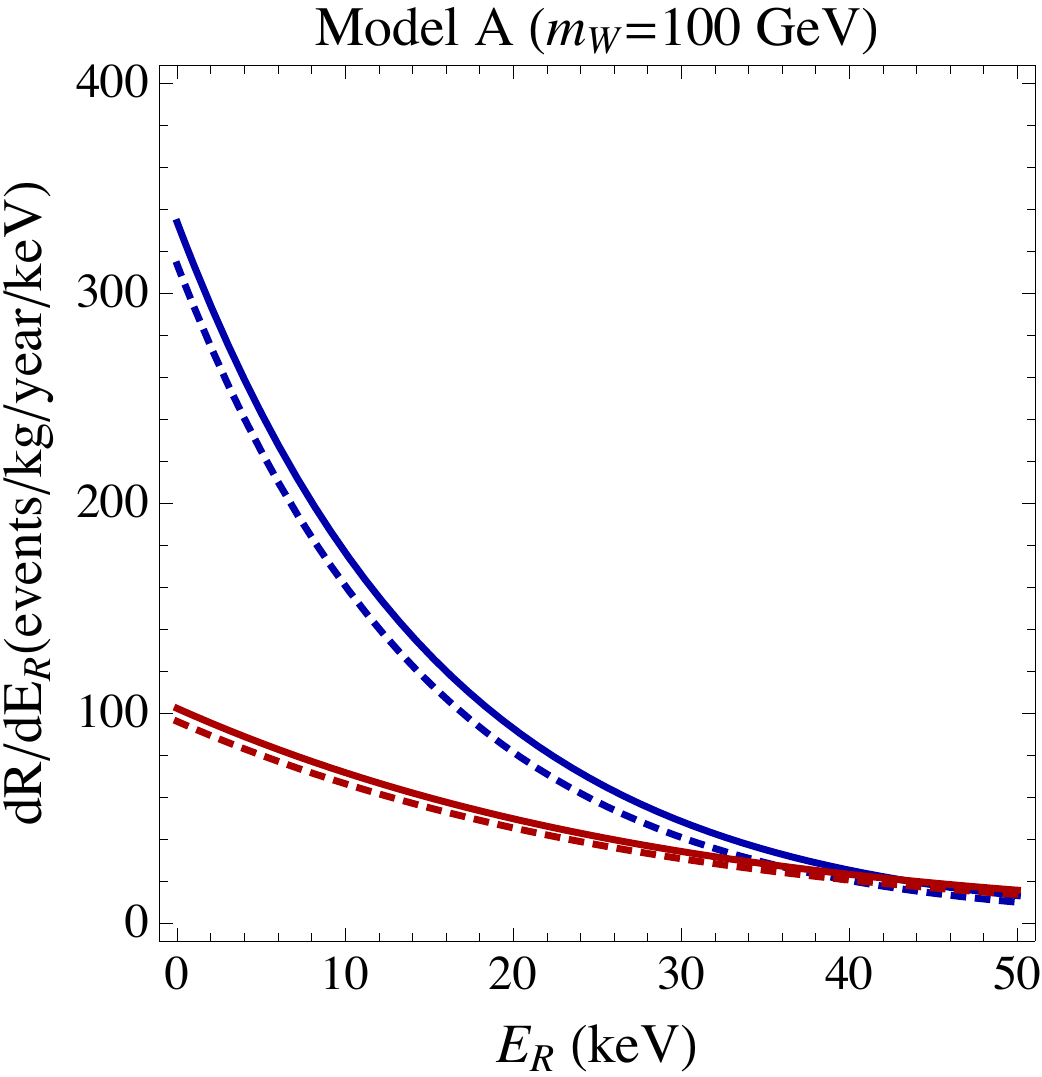}
\includegraphics[width=0.37\textwidth]{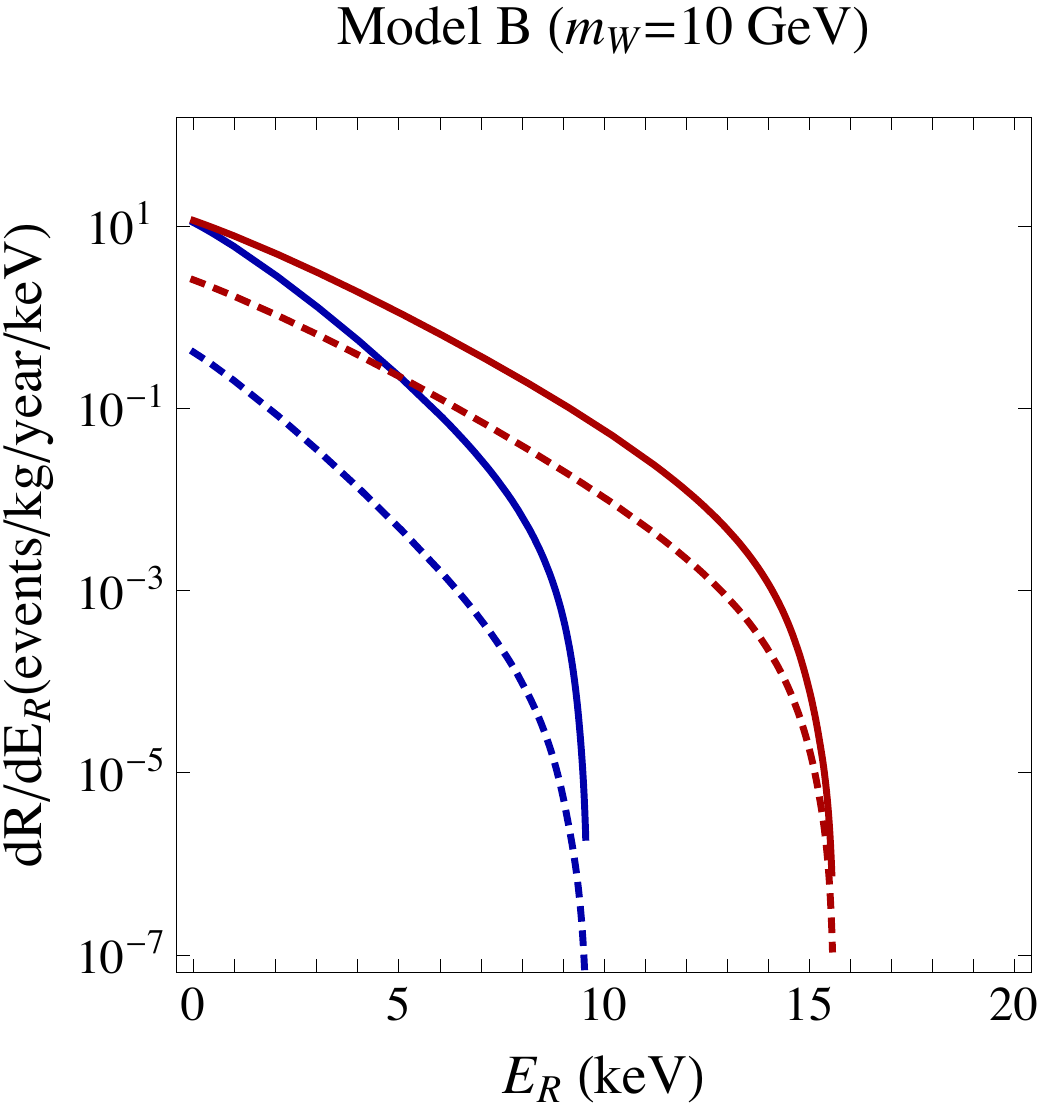}   \hspace{0.05\textwidth}   \includegraphics[width=0.36\textwidth]{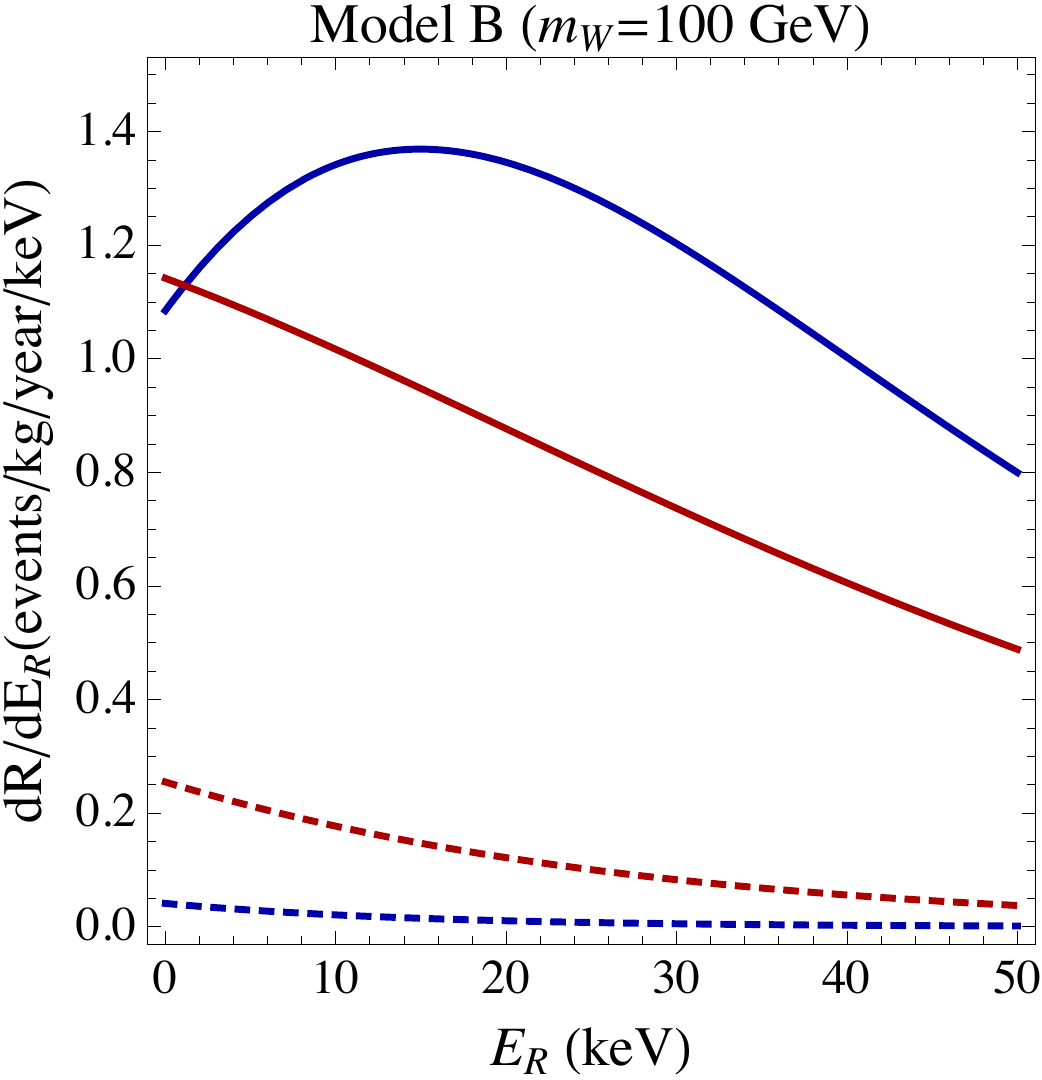}
\vspace{-.2cm}
\caption{\label{fig:pheno1v2} 
Recoil spectra for model A  ($\bar{f}_p = 0.5$~GeV,  $r =1$, $\lambda_{s,\Theta}=1$, top panels) 
and model B  ($\bar{f}_p = 0.5$~GeV,  $r = -  0.7$, $\lambda_{s,\Theta}=1$ bottom panels), 
for both Xenon (blue lines) and Germanium (red lines)  to  LO (dashed lines) and NLO (solid lines). }
\vspace{-0.5cm}
\end{center}
\end{figure}

\begin{figure}[!t]
\begin{center}
\includegraphics[width=0.4\textwidth]{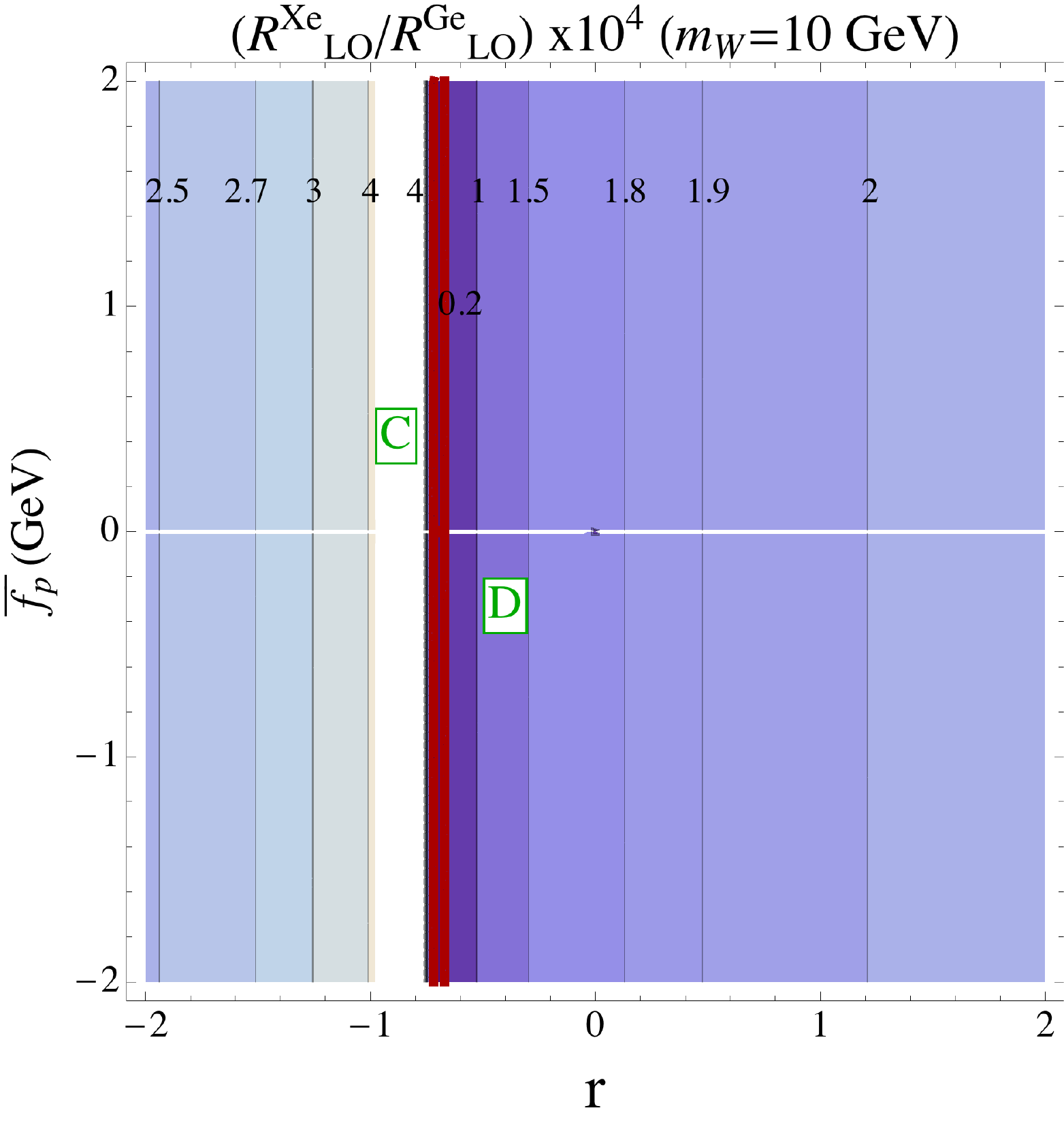}   \hspace{0.05\textwidth}   \includegraphics[width=0.4\textwidth]{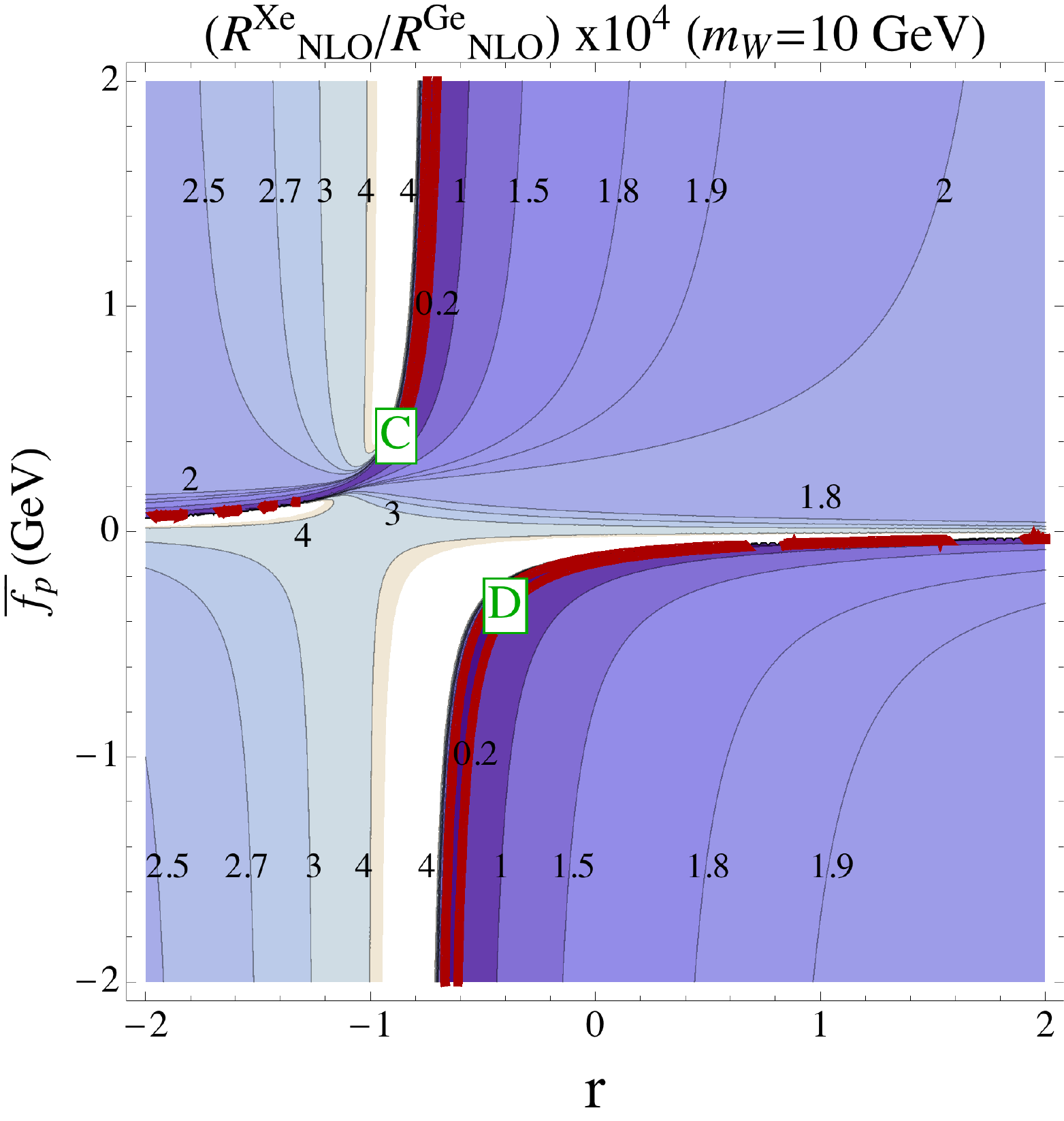}
\includegraphics[width=0.4\textwidth]{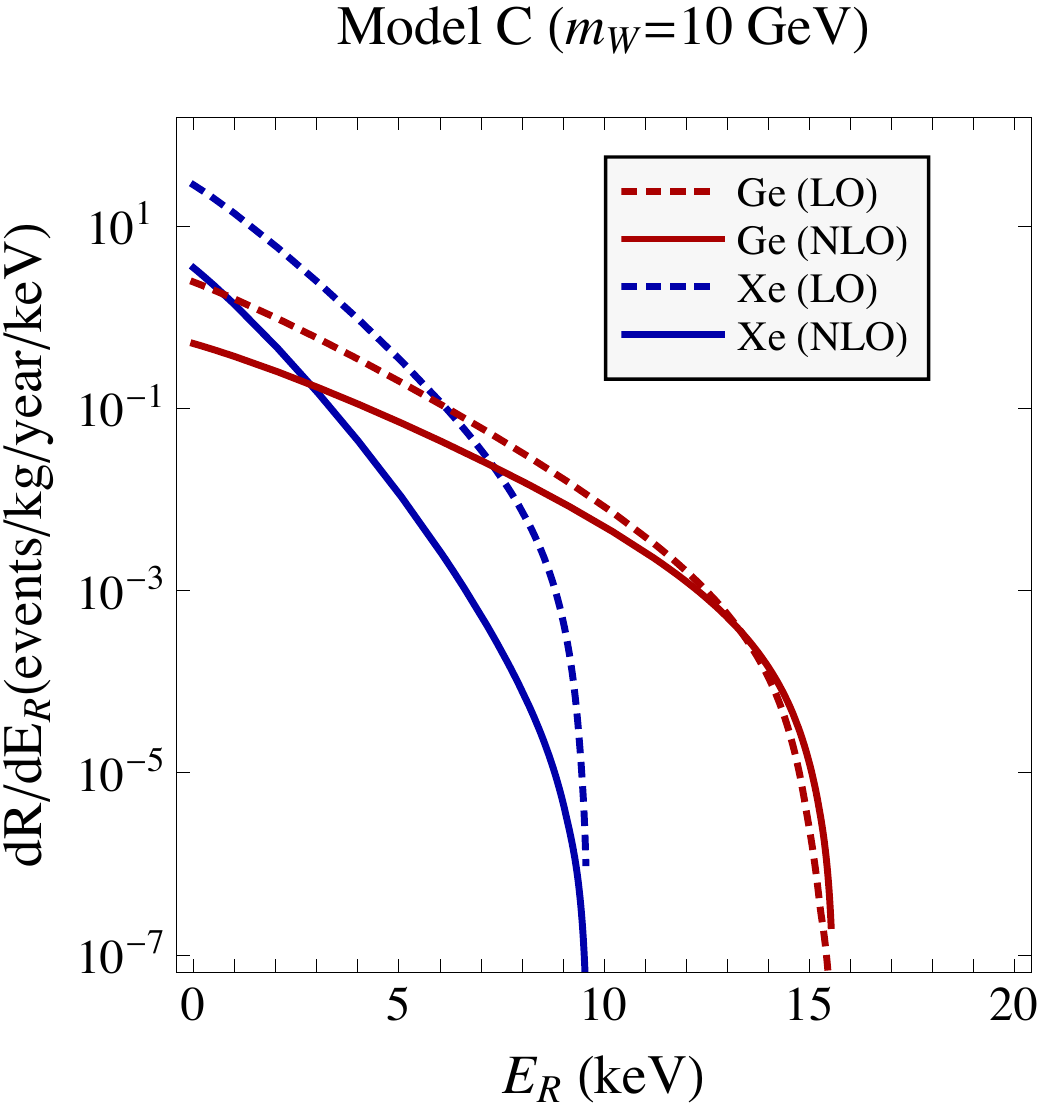}  \hspace{0.05\textwidth}    \includegraphics[width=0.4\textwidth]{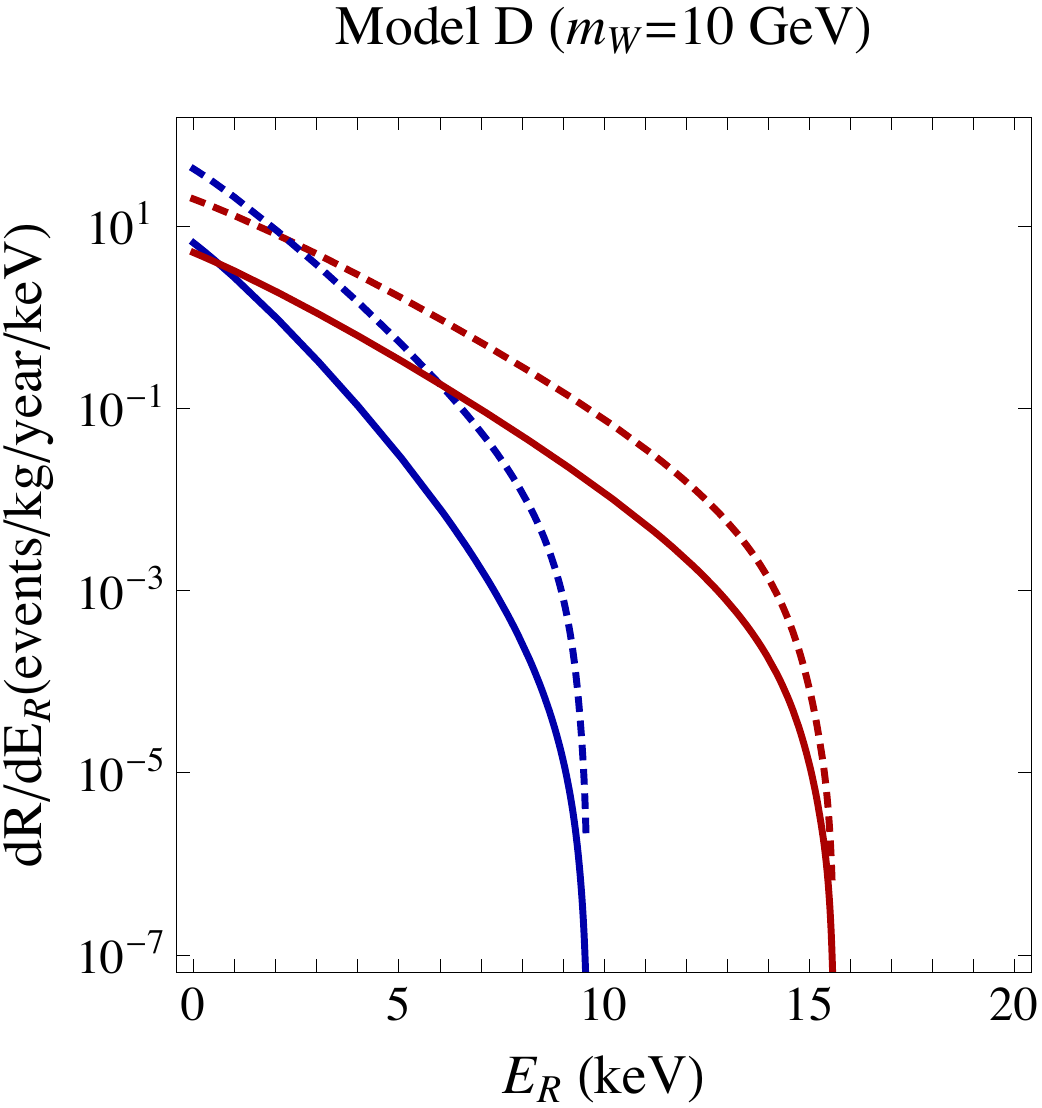}
\caption{\label{fig:pheno2} 
Top panels: contour plots of  the ratio  of Xenon and Germanium integrated rates 
$R^{\rm Xe}/R^{\rm Ge}$   on the   $(r, \bar{f}_p)$ plane,   at fixed  $\lambda_s/\lambda_\Theta=1$ 
and $m_W = 10$~GeV   to LO (left panel) and NLO (right panel). Solid red lines on top panels represent the contour lines, where $R^{\text{Xe}}_{\text{NLO}}/R^{\text{Ge}}_{\text{NLO}}=2\times10^{-5}$. Everywhere inside the solid red lines the signal in CoGent is consistent with null signal in Xenon100. 
Bottom panels:  recoil spectra for model C  ($\bar{f}_p = 0.425$~GeV,  $r =-0.885$, $\lambda_{s,\Theta}=1$) 
and model D  ($\bar{f}_p = -0.33$~GeV,  $r = -  0.4$, $\lambda_{s,\Theta}=1$), 
for both Xenon (blue lines) and Germanium (red lines)  to  LO (dashed lines) and NLO (solid lines).}
\vspace{-0.5cm}
\end{center}
\end{figure}

We illustrate  the phenomenological  implications of our  new  
WIMP-nucleus amplitude  parameterization in Figs.~\ref{fig:pheno1},  \ref{fig:pheno1v2}, 
and  \ref{fig:pheno2}: 
\begin{itemize} 
\item  In  Fig.~\ref{fig:pheno1}
we  present  contour plots  of the ratio of NLO to LO integrated rates  
$R_{\rm NLO}/R_{\rm LO}$   on the plane  
$(r, \bar{f}_p)$, 
fixing $\lambda_s/\lambda_\Theta=1$. 
We have chosen one representative target, Xenon, for which we considered  a weighted average of all 
naturally occurring isotopes and the integration region $E_R \in  [8.4, 44.6]$~keV~\cite{Aprile:2011hi}.  
We plot results for  two representative values of  the WIMP mass $m_W = 10$~GeV  (left panel)
and $m_W=100$~GeV (right panel). 
Qualitatively similar features  arise  for different choices of $\lambda_s/\lambda_\Theta \in [-50,50]$ and 
other target materials, such as Germanium.
The plots clearly display the importance of NLO corrections whenever the LO rate vanishes or is  highly suppressed, 
which happens for 
$f_p=0$ (for any finite $r$) and 
$Z + (A-Z) r = 0$  ($r \approx - 0.7$ for Xenon isotopes). 
Along these singular directions  the ratio $R_{\rm NLO}/R_{\rm LO}$  diverges or is highly enhanced. 
Moving away from these singular regions, the  ratio $R_{\rm NLO}/R_{\rm LO}$  decreases,  but corrections 
remain substantial over large regions of parameter space. 
We quantify this statement by highlighting in red the contours where  $R_{\rm NLO}/R_{\rm LO} = 2$: 
within the region enclosed by these contours the fractional corrections to the rate exceed 100\%.

\item   
In  Fig.~\ref{fig:pheno1v2}  we illustrate the impact of chiral corrections on the recoil spectra, 
for two  benchmark points in the $(r,\bar{f}_p)$ plane,   
model A where $R_{\rm NLO}/R_{\rm LO}-1\sim \mathcal{O}(10\%)$ (top panels),  
and model B  where  $R_{\rm NLO}/R_{\rm LO}$ is dramatically 
enhanced (bottom panels). 
In these plots we use $\Lambda_{\text{np}}=100\,\text{GeV}$, $v=246\,\text{GeV}$  and $\lambda_\Theta =1$  (the scaling  
 of the rate with these parameters is trivial). 
The main message  is that  while for low-mass WIMP  ($m_W = 10$~GeV, left panels) the recoil spectrum 
gets mostly  a normalization correction  with no dramatic change in the shape regardless of the value of $r$, 
for larger WIMP  masses  ($m_W = 100$~GeV, right panels) the recoil spectrum is considerably distorted when $Z + (A - Z) r \approx 0$. 
This result arises from the competition  between the linearly rising  $f_{n,p} (E_R)$  and the 
exponentially falling velocity integral $\eta (E_R, m_W, m_A)$, which cuts off at lower $E_R$  for lower $m_W$.

\item  Finally,  in Fig.~\ref{fig:pheno2},  we explore to what extent  
in our  general framework for scalar-mediated WIMP-nucleus interactions 
we can  reconcile    the tension between CoGeNT~\cite{Aalseth:2011wp}, 
which  favors a low-mass WIMP
and XENON100~\cite{Aprile:2011hi}, 
which  puts  an upper limit on the rate in this mass region. 
The tension can be quantified as follows:  
for $m_W = 10$~GeV  the standard LO  fit  with $r = f_n/f_p= 1$ implies~\cite{Kopp:2011yr}  
$\sigma_p^{\rm XENON100}  <     4 \times 10^{-43} {\rm cm}^2$  
and   $\sigma_p^{\rm CoGeNT} >   4 \times 10^{-42} {\rm cm} ^2$ (assuming large contaminations in 
CoGeNT~\cite{Kopp:2011yr}), and hence   
$\sigma_p^{\rm XENON100} / \sigma_p^{\rm CoGeNT} <  0.1$.   
In turn,  this can be converted into an upper bound on the ratio of integrated rates 
$R^{\rm Xe}/R^{\rm Ge}$ at $m_W=10$~GeV,  for any energy window.   
Using $E_R \in  [8.4, 44.6]$~keV~\cite{Aprile:2011hi} for  Xe (XENON100)  
and  $E_R \in  [2.3, 11.2]$~keV~\cite{Farina:2011pw} for Ge  (CoGeNT)   we find   
$R^{\rm Xe}/R^{\rm Ge} <  2 \times 10^{-5}$. 

In Fig.~\ref{fig:pheno2}  we show contour plots of 
$R^{\rm Xe}/R^{\rm Ge}$ 
in the $(r, \bar{f}_p)$ plane 
to LO (top left panel) and   NLO with $\lambda_s/\lambda_\Theta = 1$ (top right panel).
In these plots we also highlight in red the curves along which 
$R^{\rm Xe}/R^{\rm Ge}  =  2 \times 10^{-5}$. 
As seen from the top left panel,  assuming LO cross-sections  there is a narrow region 
around $r= -0.7$  consistent with experimental constraints. 
This is the well known regime of isospin violating dark 
matter (IVDM)~\cite{Kurylov:2003ra,Giuliani:2005my,Chang:2010yk,Feng:2011vu}.   
However, as expected from Fig.~\ref{fig:pheno1}  and explicitly  shown in  the top right panel of Fig~\ref{fig:pheno2}, 
along the $r =-0.7$  line the LO analysis cannot be trusted.  
Interestingly, our  results  show that to NLO there are still regions of parameter space 
consistent with  
$R^{\rm Xe}/R^{\rm Ge}  <  2 \times 10^{-5}$,  
which are non-trivial  deformations of the narrow band around $r \approx -0.7$. 
In these regions, the NLO corrections provide a 90\% suppression of the 
Xenon rates, i.e. $R_{\rm NLO}^{\rm Xe}/R_{\rm LO}^{\rm Xe} < 0.1$, 
again pointing to the importance of the new effects. 
We have checked that  even changing the energy-integration regions the same features emerge.
For completeness, in the bottom panels  of Fig.~\ref{fig:pheno2} we  show the recoil spectra corresponding to  
two points in parameter space   (marked as C and  D in the top panels)   consistent with 
$R^{\rm Xe}/R^{\rm Ge}  <  2 \times 10^{-5}$ to NLO.

\end{itemize}

The main features of the results presented  in Figs.~\ref{fig:pheno1},  \ref{fig:pheno1v2},   and \ref{fig:pheno2} 
are robust against changes in the  hadronic matrix elements $\sigma_{q}^{(p,n)}$ and the low-energy constant 
$\alpha = F/(F+D)$.   We have varied these inputs in the ranges  presented in Section~\ref{sect:nlo} 
and verified that the changes in Figs.~\ref{fig:pheno1}, \ref{fig:pheno1v2},  and \ref{fig:pheno2} are at the 5\% level at most. 
This  uncertainty  grows to about  20\% level when $\lambda_s/\lambda_\Theta  \gg 1$.
A related  important question for the phenomenology is: 
how robust is the one-loop ChPT  calculation of the slope of the scalar form factors? 
The dispersive analysis of Ref.~\cite{Gasser:1990ap} reveals that 
one-loop results severely under-estimate  (by more than a factor of 2)  
the slope of the iso-scalar  ($\bar{u} u + \bar{d}d$) form factor.  
We expect that   larger slopes will increase the  impact of    
recoil-energy dependent form factors  ($f_{n,p} (E_R)$), 
reinforcing the conclusions of our work. 
Therefore,   this issues deserves to be revisited in the future.

\section{Conclusions}\label{sec:conclusions}

We have applied  systematic chiral  effective theory methods to  WIMP-nucleus interactions. 
Focusing on the case of scalar-mediated WIMP-quark interactions, but otherwise in a completely model-independent 
framework,   we  have worked out the NLO corrections  to WIMP-nucleon interactions. 
A similar analysis can be done  for pseudo-scalar, vector, pseudo-vector and tensor WIMP-quark interactions.
We find that at NLO two types of effects  enter. 
First,  one loop diagrams generate 
recoil-energy dependent  corrections to the single-nucleon scalar form factors. 
The second effect involves  two-nucleon interactions with the WIMP at tree level. This generates 
a new two-body term in the WIMP-nucleus scattering amplitude. 
Our results for the modified rate formula (\ref{eq:rate1})
show that  the scalar-mediated WIMP-nucleus  cross-section cannot be parameterized 
in terms of just two quantities, namely $f_{p}$ and $f_{n}$ or 
equivalently  the WIMP-proton cross-section  $\sigma_p \propto   m_p^2  f_p^2$ and $r = f_n/f_p$. 
Two more parameters are needed for a complete description  consistent with long-distance QCD effects. 

In our model-independent scan of parameter space,  we have found  that 
the new effects  become extremely important  when  the leading order contribution
to the WIMP-nucleus amplitude  is  moderately to highly suppressed.  
We have identified the region of parameter space in which the 
fractional corrections are greater than 100\%, 
showing that it  includes the  so-called  ``isospin-violating dark matter"  regime.  
We have also  explored to what extent  the tension between CoGeNT~\cite{Aalseth:2011wp} 
and XENON100~\cite{Aprile:2011hi}, 
quantified by the ratio of integrated rates $R^{\rm Xe}/R^{\rm Ge}  <  2 \times 10^{-5}$,  
can be reconciled in our framework. 
Intriguingly,  we find  that there are regions of parameter space 
consistent with this constraint and we show that 
in these regions, the NLO corrections provide a 90\% suppression of the 
Xenon rates, i.e. $R_{\rm NLO}^{\rm Xe}/R_{\rm LO}^{\rm Xe} < 0.1$. 
Finally, we have also explored how the new corrections affect the recoil spectra, finding 
large distortions for regions in which the LO contribution is suppressed and for 
WIMP masses $m_W \geq 30$~GeV.

Both the theoretical and phenomenological analysis  presented here  should be 
regarded as only the first step of a broader  program. 
On the theory side,  we can identify several areas where future work is highly 
desirable: (i) the extension of the  ChPT analysis  beyond scalar-mediated 
WIMP-quark interactions; 
(ii)  an improved treatment of the  slope  of the 
single-nucleon scalar form factors; 
(iii)  an improved analysis of the two-nucleon matrix element, 
that goes  beyond closed shells and includes 
the dependence on the recoil energy  $E_R$. 
On the phenomenology side, we have presented a few illustrations of how our new results 
affect  direct DM detection,  without any attempt to a complete analysis.
We argue, however, that  our results call for a new model-independent  analysis of direct DM search data that 
properly includes  long-distance QCD effects,  because they can affect both the shape 
and normalization  of the recoil spectrum.

\vspace{9mm}
{\Large\bf Acknowledgements\/}\\\\
We have benefited from many helpful discussions with   Joe Carlson, Joe Ginocchio,  and Anna Hayes. 
We also thank John Donoghue and Rocco Schiavilla for discussions and Alberto Aparici for 
collaboration at  an early stage of this work. 
This work is supported by the DOE Office of Science and the LDRD program at Los Alamos.

\bibliographystyle{doiplain}
   \let\oldnewblock=\newblock
    \newcommand\dispatcholdnewblock[1]{\oldnewblock{#1}}
    \renewcommand\newblock{\spaceskip=0.3emplus0.3emminus0.2em\relax
                           \xspaceskip=0.3emplus0.6emminus0.1em\relax
                           \hskip0ptplus0.5emminus0.2em\relax
                           {\catcode`\.=\active
                           \expandafter}\dispatcholdnewblock}
\bibliography{biblio}

\end{document}